%% file: master.tex
\pdfoutput=1
\RequirePackage{ifpdf,color}
\documentclass[12pt,a4paper]{JHEP3}

\usepackage{amssymb}
\usepackage{graphicx}
\usepackage{amsmath}
\usepackage{comment}
\usepackage{mciteplus}
\usepackage{todonotes}
\presetkeys{todonotes}{inline}{}
\usepackage{booktabs}
\usepackage{placeins}

\include{declare}
\def\beq{\begin{equation}}
\def\eeq{\end{equation}}
\def\beqn{\begin{eqnarray}}	
\def\eeqn{\end{eqnarray}}

\def\as{\alpha_{\rm S}}

\def\e{{\rm e}}

\def\comix{\textsc{Comix} }
\def\sherpa{\textsc{Sherpa} }

\def\caesar{\textsc{Caesar} }
\def\powheg{\textsc{Powheg} }
\def\trace{\text{Tr}}

\def\O#1{{\cal O}(\as^{#1})}

\def\fig#1{{Fig. \ref{#1}}}
\def\eq#1{{(\ref{#1})}}
\def\nc{N_C}
\def\tf{ {\bf t} }
\def\al{\alpha}
\def\be{\beta}
\def\ga{\gamma}
\def\si{\sigma}
\def\ope{{\mathcal{O}}}
\def\T{{\bf T}}

\title{\boldmath Soft evolution of multi-jet final states}
\author{
Erik Gerwick, 
Steffen Schumann\\
II. Physikalisches Institut, 
Universit\"at G\"ottingen, 
Friedrich-Hund-Platz 1,
37077 G\"ottingen, 
Germany}
\author{Stefan H\"oche\\
SLAC National Accelerator Laboratory, Menlo Park, CA 94025, USA
}
\author{Simone Marzani\\
Center for Theoretical Physics, Massachusetts Institute of Technology, Cambridge, MA 02139, USA}

\preprint{
   SLAC-PUB-16143 \\
   MIT-CTP-4608 \\
   MCNET-14-28 \\

}

\abstract{

We present a new framework for computing resummed and matched distributions in processes 
with many hard QCD jets. The intricate color structure of soft gluon emission 
at large angles renders resummed calculations highly non-trivial in this case.  
We automate all ingredients necessary for the color evolution of the soft function at next-to-leading-logarithmic accuracy, 
namely the selection of the color bases and the projections of color operators and Born amplitudes onto those bases. 
Explicit results for all QCD processes with up to $2\to 5$ partons are given.  We also devise a new tree-level matching scheme 
for resummed calculations which exploits a quasi-local subtraction based on the Catani--Seymour dipole formalism. 
We implement both resummation and matching in the \sherpa event generator.
As a proof of concept, we compute the resummed and matched transverse-thrust distribution for hadronic collisions.
} 
\keywords{QCD Phenomenology, Jets, Hadronic Colliders}

\begin{document}

\input{inputs/intro}
\input{inputs/Sfunc}

\input{inputs/Sfunc_results}

\input{inputs/resum}

\input{inputs/summary}

\section*{Acknowledgments}
We thank Andrea Banfi for many discussions during the course of this work.
We furthermore thank Piero Ferrarese and Edgar Kellermann for useful cross checks
of some results, and Enrico Bothmann for help with some of the plotting.

This work is supported by the U.S. Department of Energy under Contract 
Number DE--AC02--76SF00515 and under cooperative research agreement Contract 
Number DE--SC00012567. The work of S.M.\ is supported by the U.S.\ National Science Foundation, under 
grant PHY--0969510, the LHC Theory Initiative. We acknowledge financial support from BMBF under 
contract 05H12MG5 and from the EU MCnetITN research network. MCnetITN is a Marie Curie Training Network 
funded under Framework Programme 7 contract PITN--GA--2012-315877.
S.H.\ thanks the Center for Future High Energy Physics at IHEP for hospitality during the final
stages of this work.

\FloatBarrier

\appendix
\input{inputs/appendix}

\bibliographystyle{JHEP}
\bibliography{journal}

\end{document}

%% file: inputs/intro.tex
\section{Introduction}\label{sec:intro}

Jets play a central role in the physics program of the CERN Large Hadron Collider (LHC). The typical minimum value for jet transverse momenta considered in LHC analyses is of the order of 20~GeV, which is more than two orders of magnitude smaller than the center-of-mass energy, resulting in a huge phase space for jet production. Events with a high jet multiplicity are therefore copiously produced at the LHC~\cite{Aad:2011tqa,*Aad:2013ysa,*Khachatryan:2014zya}.

Moreover, typical signatures of new-physics models include cascade decays of new heavy states  producing relatively hard quarks and gluons, which seed hard jets. Accurate theoretical estimates of the related QCD multi-jet backgrounds are therefore essential. This has triggered intense activity in the QCD community, resulting in more and more accurate calculations of cross sections and differential distributions for multi-jet final states.

Leading order (LO) perturbative QCD calculations for multi-jet processes can automatically be performed for large multiplicities~\cite{Mangano:2002ea,*Cafarella:2007pc,Gleisberg:2008fv}. Next-to-leading order (NLO) corrections have also reached a high level of automation~\cite{Denner:2005nn,*Ossola:2006us,*Ellis:2007br,*Binoth:2008uq,*Berger:2008sj,*Bevilacqua:2011xh,*Cullen:2011ac,*Cascioli:2011va,*Hirschi:2011pa,*Badger:2012pg,*Actis:2012qn}, and fully differential multi-jet cross sections are now available for pure QCD processes and electroweak ($W^\pm$, $Z$ and Higgs) boson production in association with up to five jets~\cite{Ita:2011wn,*Bern:2013gka,*Badger:2013yda,*Cullen:2013saa}. 

Monte Carlo parton showers~\cite{Marchesini:1983bm,*Sjostrand:1985xi,*Marchesini:1987cf}, which describe the all-order evolution of QCD partons fully exclusively, have been extended beyond the strict collinear limit~\cite{Nagy:2006kb,*Giele:2007di,*Schumann:2007mg,*Platzer:2009jq} and even beyond the $1/\nc$ approximation~\cite{Platzer:2012np,*Nagy:2012bt}. They can be merged with LO predictions for multi-jet events~\cite{Catani:2001cc,*Mangano:2001xp,*Lonnblad:2001iq,*Krauss:2002up} and matched to NLO calculations~\cite{Frixione:2002ik, Nason:2004rx,*Frixione:2007vw} for over a decade. More recently, methods for combining next-to-leading order matched predictions of varying jet multiplicity have been devised~\cite{Hoeche:2012yf,*Frederix:2012ps,*Lonnblad:2012ix}, as well as matching methods at next-to-next-to leading order (NNLO) accuracy~\cite{Hamilton:2012rf,*Hamilton:2013fea,*Hoeche:2014aia}. Dedicated Monte Carlo programs aimed at better describing jet production in the high-energy limit have also been developed~\cite{Andersen:2011hs}.

Thus, the past years have brought substantial theoretical progress in multi-jet physics, both from the viewpoint of fixed-order calculations and parton showers, as well as the matching and merging of the two approaches. 
Another important aspect of QCD phenomenology is the all-order resummation of particular classes of observables or processes, beyond the leading-logarithmic (LL) accuracy, which is typical for parton showers. Event shapes in electron-positron, electron-proton and hadron-hadron collisions have been studied for a long time (see for instance~\cite{Dasgupta:2003iq, Banfi:2010xy} and references therein) and a general framework for resumming event shapes at next-to-leading logarithmic (NLL) accuracy was developed in Refs.~\cite{Banfi:2001bz,*Banfi:2003je,*Banfi:2004nk,*Banfi:2004yd}. Very high logarithmic accuracy (N$^3$LL) was achieved using Soft Collinear Effective Theory (SCET) for particular event shapes in $e^+e^-$ collisions~\cite{Abbate:2010xh,Hoang:2014wka}. Inter-jet radiation and in particular its response to the presence of a jet veto has also received a lot of attention both from the theoretical~\cite{Oderda:1998en,*Appleby:2003sj,Forshaw:2006fk,*Forshaw:2008cq,*Forshaw:2009fz,*DuranDelgado:2011tp,Liu:2012sz} and experimental~\cite{Aad:2011jz,*ATLAS:2012al,*Aad:2014pua,*Chatrchyan:2012gwa} communities, primarily in the context of Higgs-boson studies~\cite{Banfi:2012yh,*Banfi:2012jm,*Banfi:2013eda,*Becher:2012qa,*Becher:2013xia,*Stewart:2013faa,*Boughezal:2013oha}. All-order analytical calculations have been performed recently for an increasing number of jet-substructure observables, including jet masses~\cite{Li:2012bw,*Dasgupta:2012hg,*Chien:2012ur,*Jouttenus:2013hs}, other jet shapes~\cite{Larkoski:2013eya,*Larkoski:2013paa,*Larkoski:2014uqa,*Larkoski:2014tva,*Larkoski:2014gra}, sub-jet multiplicity~\cite{Gerwick:2012fw} and grooming algorithms~\cite{Dasgupta:2013ihk,*Dasgupta:2013via,*Larkoski:2014wba}. Recently, there has also been substantial progress towards achieving NNLL accuracy in threshold resummation for dijet production~\cite{Broggio:2014hoa,*Hinderer:2014qta}.

However, to our knowledge, all phenomenological studies that used all-order resummed results have been restricted to cases with four or less hard colored partons, i.e.\ $2\to 2$ QCD scattering in hadron-hadron collisions~\cite{Dokshitzer:2005ek,*Dokshitzer:2005ig,Kidonakis:1998nf,Bonciani:2003nt}~\footnote{Refs.~\cite{Forshaw:2006fk,*Forshaw:2008cq,*Forshaw:2009fz,*DuranDelgado:2011tp} considered the resummation of $2\to3$ scattering processes, in the limit where one of the final-state partons was a soft gluon.}. The reason for this deficiency in comparison to the enormous progress in fixed-order calculations is purely technical. While logarithmic terms associated to collinear emissions have a simple color structure, i.e.\ the Casimir operator of the jet under consideration, the color structure of soft-gluon emissions at large angles is more complex and, in particular, has a non-trivial matrix structure for $n\ge 4$ partons. Nevertheless, resummed calculations can in principle be written for an arbitrary number of hard colored legs, using, for instance, the formalism of Refs.~\cite{Dokshitzer:2005ek,*Dokshitzer:2005ig,Catani:1985xt,Catani:1996jh,*Catani:1996vz,Bassetto:1984ik}. In order to perform an actual calculation, one then needs to define a suitable color basis for each partonic subprocess, and consequently find the matrix representation of all color insertions. The dimensionality of color bases rapidly increases with the number of legs. Algorithms to define them have been discussed in the literature, e.g.~\cite{Sjodahl:2008fz, Sjodahl:2009wx,Sjodahl:2012nk,Keppeler:2012ih}. However, when making use of a non-orthogonal basis, the efficient inversion of the matrix representing the color metric can pose a severe problem. In addition, the underlying Born matrix elements for the hard process must be decomposed in the chosen basis. One would clearly like to automate all these steps. 

The main purpose of this study is to overcome these technical difficulties and provide a tool to perform soft-gluon resummation at NLL accuracy for processes with, in principle, arbitrarily many hard legs. In practice we have considered all contributions for up to $2\to 5$ processes. As detailed in Sec.~\ref{sec:resum}, we achieve this by writing the resummed exponent in a suitable color basis and by decomposing the Born amplitudes using modified color-dressed recursive relations~\cite{Duhr:2006iq}, as implemented in the \comix matrix-element generator~\cite{Gleisberg:2008fv}, that is part of the \sherpa framework \cite{Gleisberg:2003xi,*Gleisberg:2008ta}. 

In this paper, we also address the issue of matching the resummation to fixed-order calculations. In Sec.~\ref{sec:resummation} we develop an automated LO matching scheme which makes use of modified dipole subtraction~\cite{Catani:1996jh,*Catani:1996vz}. It circumvents the explicit expansion of the resummation formulae to a large extent and provides a quasi-local cancellation of the logarithmic contributions. 
We use the resummation of the transverse thrust in hadronic collisions as a first example to study the performance of our method.
We finally summarize our work and indicate future directions in Sec.~\ref{sec:conclusions}.

%% file: inputs/Sfunc.tex
\section{The soft function and its anomalous dimension}\label{sec:resum}

The main aim of this work is to define and implement NLL resummation for processes with an arbitrary number of hard partons. 
Despite the computational difficulties arising from the non-trivial color structure in soft-gluon radiation, one can formally write all-order resummed expressions in terms of abstract color operators~\cite{Dokshitzer:2005ek,*Dokshitzer:2005ig,Catani:1985xt,Catani:1996jh,*Catani:1996vz, Bassetto:1984ik}, which are then valid for an arbitrary number of hard legs.

The quantity we are interested in is the NLL ``soft function''~\cite{Dokshitzer:2005ek,*Dokshitzer:2005ig,Kidonakis:1998nf}
\begin{equation}\label{eq:logS-basis-ind}
 \mathcal{S}(\xi)= \frac{ \langle m_0 | e^{- \frac{\xi}{2} \mathbf{\Gamma}^{\dagger}}e^{- \frac{\xi}{2} \mathbf{\Gamma}} |m_0 \rangle}{ \langle m_0 | m_0 \rangle}.
\end{equation}
In the above equation, $|m_0 \rangle$ denotes a vector in color space representing the Born amplitude,
such that the color-summed squared matrix element is $|{\cal{M}}_0|^2= \langle m_0 | m_0 \rangle$. Therefore, Eq.~(\ref{eq:logS-basis-ind}) describes the soft gluon evolution of the Born amplitude from the hard scale of the process down to the low scale, set by the observable under consideration, thus resumming to all orders the logarithmic contributions encoded in the evolution variable $\xi$. Note that the soft function defined here and used throughout this paper does not contain any collinear logarithms and the evolution variable $\xi$, the precise functional form of which may depend on the observable at hand, is single-logarithmic. This is in contrast to alternative definitions also common in the literature. Moreover, to NLL considered here, the soft function depends on the strong coupling only through the variable $\xi$. 

The soft function in Eq.~(\ref{eq:logS-basis-ind}) is defined in terms of the central object in 
our study: the soft anomalous dimension $\mathbf{\Gamma}$. Although much of the computational technology developed here can be applied to a variety of observables, in order to keep the presentation simple, we focus our discussion on global event shapes~\footnote{A general framework for resumming such observables has been developed in the context 
of the program \caesar~\cite{Banfi:2001bz,*Banfi:2003je,*Banfi:2004nk,*Banfi:2004yd}. Within this method, observables defined on Born configurations with an arbitrary number of hard partons can in principle be considered.
More details will be given in Sec.~\ref{sec:resum_with_caesar} and App.~\ref{app:caesar}.}.
For this class of observable, $\mathbf{\Gamma}$ 
can be written as
\begin{equation}\label{eq:gamma-ind}
\mathbf{\Gamma}= - 2 \sum_{i<j}  \T_i\cdot \T_j \,
\ln \frac{Q_{ij}}{Q_{12}}+   i \pi \sum_{i,j= II, FF}  \T_i\cdot \T_j \,.
\end{equation}
The first sum runs over all possible colored dipoles, with $Q_{ij}$ the respective invariant mass, i.e. 
\begin{equation}\label{eq:dipole_inv}
Q_{ij}^2=2\, p_i\cdot p_j .
\end{equation}
The second sum in Eq.~\eq{eq:gamma-ind} is over the Coulomb (or Glauber) contributions between final--final (FF)
and initial--initial (II) parton pairs. Note that the non-commutativity of $\mathbf{\Gamma}$ and $\mathbf{\Gamma}^{\dagger}$ prevents us from recombining the exponentials in Eq.~(\ref{eq:logS-basis-ind}) and leads to a physical effects from the Coulomb phase.

In order to make contact with the existing literature, we can evaluate 
Eq.~(\ref{eq:gamma-ind}) for the special case of $2\to 2$ scattering of massless partons. In this case, 
$Q_{12}=Q_{34}=\sqrt{s}$, $Q_{13}=Q_{24}=\sqrt{-t}$ and $Q_{14}=Q_{23}=\sqrt{-u}$, and the soft anomalous dimension 
becomes (see e.g.~\cite{Banfi:2001bz,*Banfi:2003je,*Banfi:2004nk,*Banfi:2004yd}) 
\begin{align}\label{eq:gamma-ind-2to2}
\mathbf{\Gamma}&=- \left(\T_1 \cdot \T_3+\T_2 \cdot \T_4 \right) T- \left(\T_1 \cdot \T_4+\T_2 \cdot \T_3 \right) U,
\end{align}
where we have employed color conservation, i.e.
\begin{align}\label{eq:colcons}
\left(\sum\limits_{i=1}^4\T_i \right) | m_0 \rangle = 0\,, 
\end{align} 
%
%\begin{align}\label{eq:colcons}
%| m_0 \rangle \langle m_0 | = 0\,, 
%\end{align} 
introduced the compact notation
\begin{align}
 T= \ln \frac{-t}{s}+ i \pi\quad\text{and}\quad U= \ln \frac{-u}{s}+ i \pi,
\end{align}
and dropped all contributions from abelian phases because they do not contribute to any cross 
sections.

Aiming for an automated evaluation of Eq.~(\ref{eq:logS-basis-ind}) for arbitrary processes, there are 
essentially three problems which need to be addressed:
\begin{itemize}
\item the color-basis definition and computation of the metric,
\item the computation of the color operators $\T_i \cdot \T_j$, in the considered basis,
\item the decomposition of the amplitude $|m_0\rangle$ in the considered basis.
\end{itemize}

The construction and implementation of an algorithm addressing all three items represents the core of this paper.
This problem is closely related to the color decomposition of QCD amplitudes~\cite{Mangano:1990by}, which is typically written in the form
\begin{equation}\label{eq:color_decomposition}
  \mathcal{M}_0(1,{\alpha_1};\ldots;n,{\alpha_n})=\sum_{i}C^{(i)}(\alpha_1,\ldots,\alpha_n)\,m_0^{(i)}(1,\ldots,n)\;.
\end{equation}
Here, $\mathcal{M}_0$ is the full amplitude for a set of external particles $1\ldots n$
with color assignments $\alpha_1\ldots\alpha_n$. The $C^{(i)}$ are color coefficients, and the $m_0^{(i)}$ 
are color-ordered partial amplitudes. The index $i$ labels the color orderings contributing to the color assignment.
While the number of orderings and the related color coefficients change with the color basis~\cite{DelDuca:1999ha,*DelDuca:1999rs,Maltoni:2002mq}, 
the partial amplitudes are unique, gauge-invariant objects depending only on the particle momenta. 
They are given by sums of planar diagrams computed in the large-$N_C$ limit~\cite{tHooft:1973jz}. 
One may consider Eq.~\eqref{eq:color_decomposition} the projection of the Born amplitude 
onto a given color-basis element, $\mathcal{M}_0(\alpha)=\langle c_\alpha|m_0\rangle$. 
This will be discussed in more detail in the following.

\subsection{Non-orthogonal color bases}\label{sec:color-flow}

We first define our notation for color bases. As we are 
going to work with bases which are not necessarily orthogonal (for a discussion 
about this topic see also Refs.~\cite{Schofield:2011zi,Platzer:2013fha,Schofield:2013aa}),
we start by defining basis vectors $|c_\alpha\rangle$ and introduce the 
(non-diagonal) color metric, and its inverse 
\begin{equation} 
\langle c_\alpha | c_\beta \rangle = c_{\alpha\beta} \ne \delta_{\alpha\beta} \quad \quad
c^{\alpha\beta} = (c_{\alpha\beta})^{-1}\,.
\label{north_bas}
\end{equation}
Note that $c_{\alpha \gamma}c^{\gamma \beta} = \delta_{\alpha}^{\;\;\beta}$ by construction.  
The basis vectors $|c_\alpha\rangle$ span a complete (possibly over-complete) set of elements 
which we leave undetermined for the moment.  We will adopt the convention of 
referring to $c^{\alpha\beta}$ as the inverse metric.

Let us consider a general tensor $H^{\be\ga}$ expressed in the 
non-orthogonal $c$-basis.  Color invariants are computed by contracting with the metric, 
and we define in particular the color trace as ${\rm Tr}(cH)=c_{\al\be} H^{\al\be}$.  
Indices between the $c$-basis and its dual are raised and lowered with 
the metric. Tensors transforming with mixed indices are interpreted as 
$ H^{\;\;\be}_{\al} \equiv   H_{\al\ga} c^{\ga\be}$.  

The soft function from Eq.~(\ref{eq:logS-basis-ind}) written 
in matrix notation reads
\begin{alignat}{5}
\mathcal{S}(\xi)\; \; = \;\; 
\frac{{\rm Tr}\left( H \e^{-\frac{\xi}{2} \Gamma^\dagger} c \,\e^{-\frac{\xi}{2} \Gamma}\right)}{{\rm Tr}\left( c H \right)}
%\;\; = \;\;
 %\frac{ c_{\al\be} H^{ \ga \si}  \, 
%\mathcal{\bar{G}}_{\ga}^{\;\;\be} \; \mathcal{G}^{\al}_{\;\;\si}}{c_{\al\be} H^{\al\be}} 
\;\; = \;\;
 \frac{ c_{\al\be} H^{ \ga \si}  
\mathcal{G}^\dagger_{\ga \rho} c^{\rho \be} c^{\al \delta} \mathcal{G}_{\delta \si}}{c_{\al\be} H^{\al\be}}, 
\label{soften}
\end{alignat}
where $c_{\alpha\beta} H^{\alpha\beta}=\langle m_0 | m_0 \rangle$ now represents the 
color-summed Born matrix element squared.  The matrix $\mathcal{G}$ is the exponential of the 
soft anomalous dimension matrix, which due to the non-orthogonal nature of 
the $c$-basis, takes the form
\begin{equation}
\mathcal{G}_{\al\be} (\xi)
%\; \; = \;\;  c_{\al \gamma}\; \mathcal{G}_{\;\;\be}^{\gamma} (\xi)
\;\; = \;\;
c_{\al \gamma}
\exp\left 
(-\frac{\xi}{2} \,  \Gamma_{\;\;\be}^{\gamma}\right)
\;\; = \;\;
c_{\al \gamma}
\exp\left 
(-\frac{\xi}{2} \, c^{\gamma \delta} \, \Gamma_{\delta\be}\right).
\label{eq:exp}
\end{equation}
A significant amount of recent work has focused 
on improving the basis construction, with certain advantages and 
disadvantages for each approach.  In \cite{Sjodahl:2009wx}, 
a complete trace basis was discussed which followed from  
combining the connected fundamental representation color tensors 
appearing in the tree-level hard matrix element with the disconnected 
color structure required by soft-gluon exchange.  The construction of 
this basis for an arbitrary process was automated in \cite{Sjodahl:2012nk}.  
In \cite{Keppeler:2012ih} a general orthonormal basis was constructed, 
which was shown to be minimal in elements for a given process.

In this work we follow a different approach. Instead of constructing new
optimized color bases, we rely on existing ones and circumvent the problem
of over-completeness in an automated fashion by extending the dimensionality
of color space. This method, and the alternative approach of dimensional
reduction, will be discussed in more detail in Sec.~\ref{3path}.
In order to select the color bases to start with, 
we use the following guiding principles:
\begin{enumerate}
\item\label{req_minimal_partial_count}
{\bf Minimal partial-amplitude count:}  
The components $\mathcal{M}(\alpha)=\langle c_\alpha|m_0\rangle$
should depend on as few partial amplitudes as possible.

\item\label{req_physical_states}
{\bf Physical color states:}
The basis vectors should represent physical color states.
This disqualifies bases containing singlet gluons, for example. 

\item\label{req_invertibility}
{\bf Minimality of the basis:} 
Although we will also use over-complete bases, we require the dimension 
of the basis to be as low as possible.

\end{enumerate} 
The trace basis~\cite{Mangano:1990by} for processes with quarks 
and the adjoint basis \cite{DelDuca:1999ha,*DelDuca:1999rs} for processes with 
only gluons satisfy our guiding principles, and we choose to implement them. However, 
subtleties arise because these bases can be over-complete.  In this respect, 
we note that exponentiation via Eq.~\eq{eq:exp}
requires the computation of the inverse color metric 
$c^{\alpha\beta}$. Thus, $c^{\alpha\beta}$ must be non-singular for general 
$\nc$.  At $\nc =3$ it may be singular if the corresponding 
metric $c_{\alpha\beta}$ contains representations with 
weight proportional to $\nc - 3$.  In this case the 
inversion may be computed with $\nc = 3 + \epsilon$ colors. More on this 
issue appears in Sec.~\ref{3path}.

\subsubsection*{Processes including quarks}
The complete basis for processes including quarks follows from color 
connecting all same flavor quark lines while attaching gluons in the form of
fundamental-representation matrices. For example, in the case of a single quark pair 
the decomposition at tree-level is~\cite{Mangano:1990by}
\begin{equation}\label{eq:cc_quark}
  \mathcal{M}_0(1,{i_1};2,{a_2};\ldots;n,{j_n})=\sum_{\sigma\in P(n-2)}(T^{a_{\sigma_2}}\ldots T^{a_{\sigma_{n-1}}})_{i_1}^{j_n}\,m_0(1,\sigma_2,\ldots,\sigma_{n-1},n)\;.
\end{equation}
The sum runs over all $(n-2)!$ permutations of the particle labels $2\ldots n-1$, which represent the gluons.
Decompositions for processes with multiple quark lines are qualitatively similar and can be found in the literature.
In the general case, the decomposition includes disconnected quark lines, arising from soft 
gluon exchange, and disconnected gluon lines, which appear for processes with 2 or more gluons.
Similar terms appear at higher loops in fixed-order calculations.

An important simplification is that any basis 
with the same number of $q \bar{q}$ pairs (taking flavor labels 
as all incoming) and gluons is the same, modulo crossings.  This suggests  
that for a given set of particle flavors, the resummation may be carried out
for a fixed flavor ordering.  In practice we implement 
this by always computing ${\bf \Gamma}$ in the same 
flavor arrangement $\{q , \bar{q}, g\}$ so that the first 
sum in Eq.~\eq{eq:gamma-ind} is always in order.

We keep track of the map to the physical process by labeling incoming and 
outgoing for the purpose of assigning the Coulomb phase. This means that 
the matrices $\T_i \,\cdot \, \T_j$ only need to be 
computed once for all processes involving the same number of quarks 
and gluons. 

\subsubsection*{Purely gluonic processes}
There are multiple options of dealing with purely gluonic processes.
The first and oldest of them is the trace basis, described in 
Ref.~\cite{Mangano:1990by}. The color decomposition of tree-level amplitudes reads
\begin{equation}\label{eq:cc_gluon_trace}
  \mathcal{M}_0(1,{a_1};\ldots;n,{a_n})=\sum_{\sigma\in P(n-1)}{\rm Tr}(T^{a_1}T^{a_{\sigma_2}}\ldots T^{a_{\sigma_n}})\,m_0(1,\sigma_2,\ldots,\sigma_n)\;.
\end{equation}
The sum runs over all $(n-1)!$ permutations of the particle labels $2\ldots n$.
In the context of resummation, we must add the non-vanishing color disconnected 
components containing multiple gluon traces, which also appear at higher loops
in fixed-order calculations.

A subtlety arises due to the reflection symmetry of the partial amplitudes,
$m_0(1,2,3,\ldots,n) = m_0(1,n,\ldots,3,2)$, which holds for the corresponding 
soft gluon evolved amplitudes as well. The basis elements corresponding to 
permutation $123 \ldots n$ and $n \ldots 321$ can be combined due to this symmetry, so that 
the number of connected basis elements for general $\nc$ is reduced by a factor two. 

The adjoint ($f$-) basis \cite{DelDuca:1999ha,*DelDuca:1999rs} corresponds to the remaining
basis vectors after applying the Kleiss-Kuijf relations~\cite{Kleiss:1988ne}.
Equation~\eqref{eq:cc_gluon_trace} reduces to
\begin{equation}\label{eq:cc_gluon_adjoint}
  \mathcal{M}_0(1,{a_1};\ldots;n,{a_n})=\sum_{\sigma\in P(n-2)}(F^{a_{\sigma_2}}\ldots F^{a_{\sigma_{n-1}}})_{a_1a_n}\,m_0(1,\sigma_2,\ldots,\sigma_{n-1},n)\;,
\end{equation}
where the sum runs over only $(n-2)!$ permutations, corresponding to the new basis elements. 
As with the trace basis, we add the disconnected components, which starting at 6 gluons 
may also feature $4$ gluons connected via adjoint tensors. 

\subsection[Elimination of $\nc = 3$ pathologies]{Elimination of \boldmath{$\nc = 3$} pathologies}
\label{3path}
Although advantageous from many points of view, both the trace and adjoint bases for high-multiplicity processes turn out to be over-complete. As a consequence, the matrices representing the corresponding color metric, defined as in Eq.~\eq{north_bas}, have null eigenvalues at $\nc=3$.
However, the fact that the inverse metric at $\nc = 3 $ 
is often singular is an artefact of calculating $c^{\alpha\beta}$ 
and $\mathbf{\Gamma}$ separately, since maintaining the full $\nc$ 
dependence the resulting $\mathcal{S}$-function is always 
finite.  Keeping the $\nc$ dependence explicit 
becomes computationally impractical for large multiplicity.  Here we outline 
two strategies to overcome these limitations.  

\subsubsection*{Dimensional Reduction} 

The simplest solution is to reduce the size of the color basis, in particular, if 
we bear in mind the freedom to reparameterize the  basis elements with no 
tree-level Born contribution.  These components only enter $\mathcal{S}$ through 
contractions with the inverse metric and can therefore be reshuffled for convenience.  

More precisely, for a basis with $m$ Born proportional 
and $n-m$ non-Born elements $\{c_0, \cdots ,\, c_{m-1} , \, c_m , \, c_{m+1}, 
\cdots ,\, c_{n-1},  \, c_n  \} $,  we examine the situation where there is a 
single zero eigenvalue at $\nc =3$ in the color metric.  In other words, 
the basis decomposes into $n-1$ non-vanishing irreducible representations. 
A simple procedure for reducing the color space then corresponds to the 
new basis $\{c_0, \cdots ,\, c_{m-1} , \, c_m+c_n , \, c_{m+1}+c_n, \cdots 
,\, c_{n-1}+c_n \} $, where we normalise new elements 
accordingly.

While for simpler processes this procedure is straight-forward (see 
section~\ref{sec:gggg_ex}), for the general case it is hard to automate,
and therefore we choose a different approach.
 
\subsubsection*{Numerical Inversion with \boldmath{$\nc = 3 + \epsilon$}}

We adopt a solution which avoids adjusting the dimensionality of the 
basis, and therefore requires no \emph{a priori} group theory knowledge on 
the color decomposition of a given process.  This is the simplest 
solution practically, though there is clearly an efficiency loss due to carrying 
through non-contributing color directions.

We state the necessary claims here while proofs may be found in App.~\ref{sec:NCeps_proof}.  
First, we note that the metric is always invertible for $\nc = 3+\epsilon$ with 
$\epsilon > 0$.  We can separate the singular from the regular part of 
the inverse color metric as
\begin{equation}
c^{\al\be}_{3+\epsilon} = c^{\al\be}_R + \frac{1}{\epsilon} \tilde{c}^{\,\al\be}\, .
\end{equation}
The singular part of the inverse metric is in the null-space of all 
color products evaluated at $\nc = 3$
\begin{equation}
\tilde{c}^{\,\al\be} (\T_i \cdot \T_j)_{\beta \gamma} = {\bf{0}}^\al_{\;\;\gamma}\,,
\label{key_other}
\end{equation} 
which guarantees that
\begin{alignat}{5}
 \mathcal{S}(\xi)_{\nc =3 + \epsilon}  = \mathcal{S}(\xi)_{\nc = 3} +  \ope{(\epsilon)}. 
 \label{keyr1}
\end{alignat} 
We find that the error introduced in the resummation is $\ope(\epsilon)$ which 
may be taken sufficiently (arbitrarily) small in practice (theory).

%% file: inputs/Sfunc_results.tex
\subsection{Computation of the hard matrix}
\begin{figure}
  \begin{center}
    \includegraphics[scale=0.45]{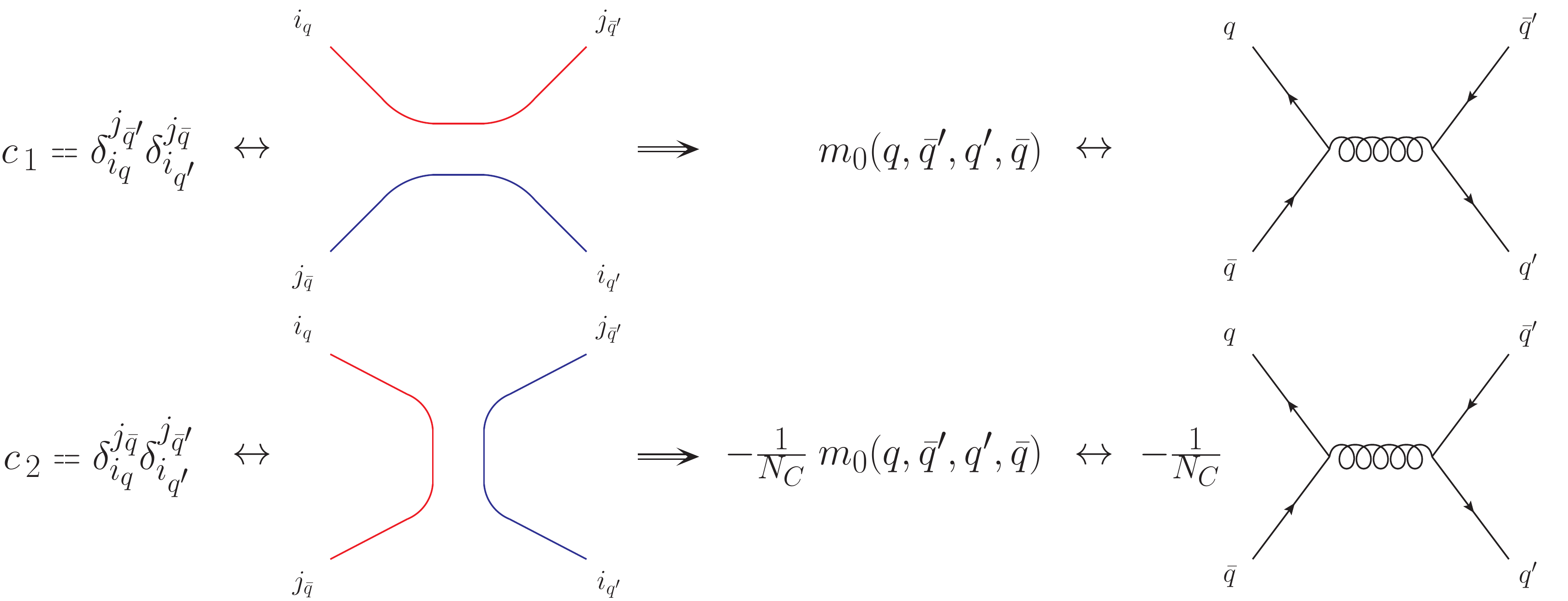}
  \end{center}
  \caption{Sketch of color basis vectors and their corresponding projections
    of Born matrix elements for $q\bar{q}\to q'\bar{q}'$ scattering.
    All flavors in the figure are taken as outgoing.
    \label{fig:hard_schannel}}
\end{figure}
\begin{figure}
  \begin{center}
    \includegraphics[scale=0.45]{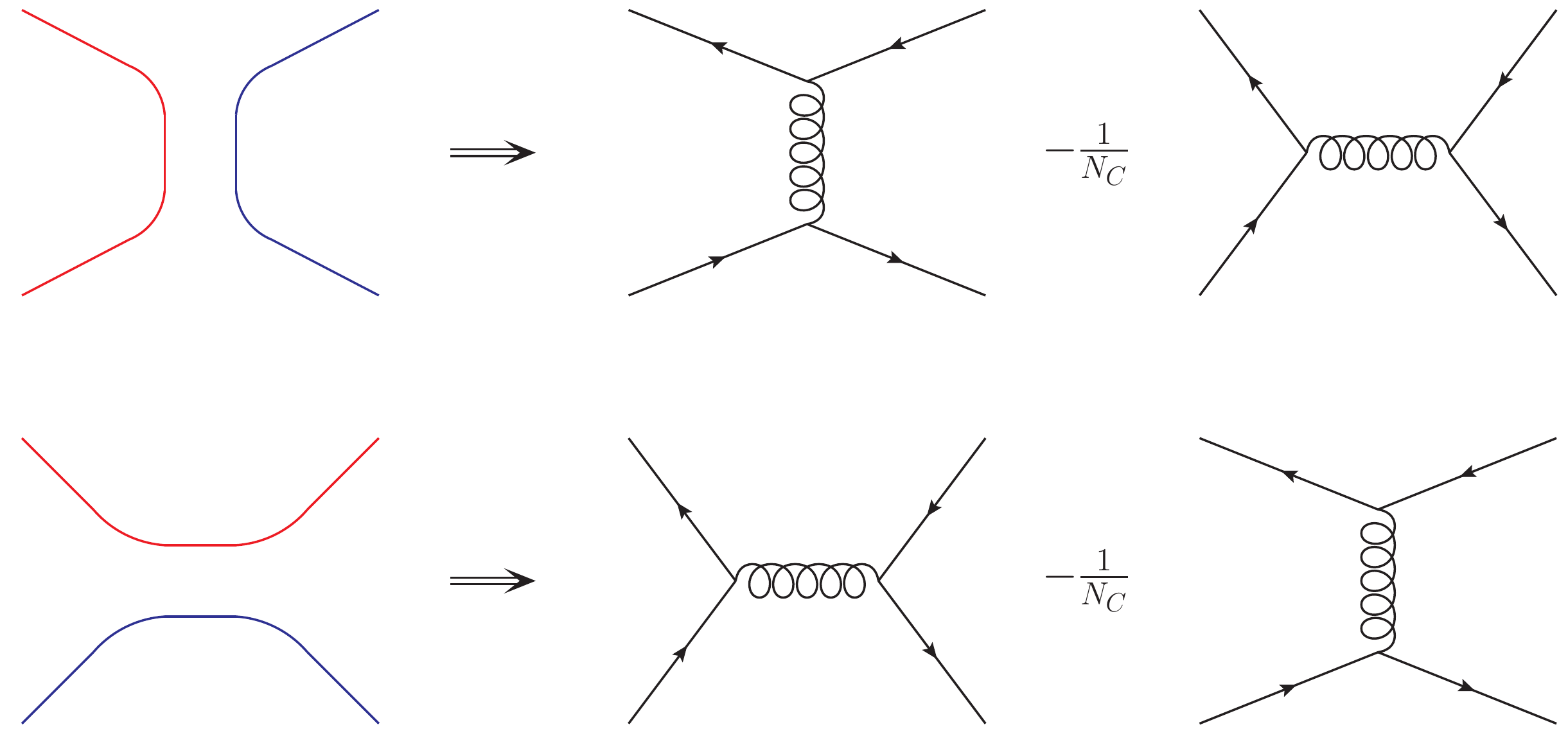}
  \end{center}
  \caption{Sketch of color basis vectors and their corresponding projections
    of Born matrix elements for $q\bar{q}\to q\bar{q}$ scattering. In comparison 
    to Fig.~\protect\ref{fig:hard_schannel}, there is both an s- and a t-channel diagram,
    both of which contribute to each projection with different weight.
    \label{fig:hard_stchannel}}
\end{figure}

A key ingredient for the computation of the soft function Eq.~(\ref{soften}) is the hard matrix, which is formed by projections of the Born amplitudes 
onto color basis vectors, $H^{\alpha\beta}=\langle m_0|c_\alpha\rangle\langle c_\beta|m_0\rangle$.
Consider, for instance, the trivial case of $q\bar{q}\to q'\bar{q}'$ scattering, where $q$ and $q'$
represent two different quark flavors. The Born matrix element factorizes into a purely
kinematical part, which stems from the $s$-channel diagram squared, and color coefficients
defining the actual matrix structure. This is shown in Fig.~\ref{fig:hard_schannel}. 
However, in any non-trivial case, multiple diagrams appear, 
which contribute differently to the different matrix elements, such that the hard matrix
has a non-trivial dependence on the Born kinematics. In particular, same-flavor quark processes 
like $q\bar{q}\to q\bar{q}$ scattering have partial amplitudes where both $s$- and the $t$-channel 
diagrams contribute because of the $1/\nc$ suppressed term in the Fierz identity. This is sketched in Fig.~\ref{fig:hard_stchannel}.
Automating the computation of $H^{\alpha\beta}$ requires an algorithm that allows us 
to easily access these partial amplitudes.

We solve this problem with the help of \comix \cite{Gleisberg:2008fv}, a matrix-element generator 
that computes multi-parton amplitudes using color-dressed recursive relations~\cite{Duhr:2006iq}.
\comix is part of the \sherpa framework \cite{Gleisberg:2003xi,*Gleisberg:2008ta}.
As \comix allows us to define a color configuration in the large-$\nc$ limit, it is trivial 
to obtain color-ordered partial amplitudes. However, these are not necessarily sufficient
to compute the entries of the hard matrix directly.

Take for example $q\bar{q}\to q\bar{q}$ scattering, as depicted in Fig.~\ref{fig:hard_stchannel}.
The two amplitudes needed for the hard matrix are shown schematically on the first and the second line.
To compute them individually, we can use a colorful matrix element that is projected onto the correct 
set of diagrams by selecting external colors appropriately. Using the color-dressed Feynman rules 
from~\cite{Duhr:2006iq}, the amplitudes on the right-hand side, including their prefactors, 
are generated by choosing the colors on the left-hand side. If the number of colors is fixed 
to three, this leads to problems for amplitudes with more than three fundamental color indices,
as non-planar diagrams start to appear. These are removed by working at $\nc\to\infty$. 
Taking this limit, however, would eliminate the second diagram on the first line and 
the first diagram on the second line, because the gluon propagator does not have a $1/\nc$ 
contribution. The problem is solved by keeping this term when taking the limit. This modification is implemented at the vertex level by changing the color-dressed Feynman rules 
such that $U(1)$ gluons couple to quark lines also in the large-$\nc$ limit, while evaluating 
the corresponding $1/\nc$ term in the Fierz identity with $\nc=3$.

We note that it is possible to add the relevant one-loop partial amplitudes and
extend this algorithm beyond the tree level. This will provide the hard matrix one
order higher, which is needed in order to achieve higher logarithmic accuracy in the
resummation.

\subsection{Validation against multi-parton matrix elements}

In order to check the construction of the color metric for the 
employed bases and the correctness of the corresponding decomposition 
of the hard matrix for multi-parton amplitudes, we compare 
our results against exact real-emission matrix elements 
considering soft but non-collinear kinematics for the emitted
gluon. Starting from an $n$-parton state with momenta $p_1,\ldots, p_n$
we assume the emitted gluon to carry additional momentum $p_s$, with $|p_s|=k_s$.
We choose a particular kinematic configuration, where the final-state momenta 
resemble a circle in the transverse plane, i.e.,
\begin{alignat}{5}\label{eq:hard_soft_fac_test}
&p_1 = E(1,0,0,1)\,,\notag\\
&p_2 = E(1,0,0,-1)\,, \notag \\
&p_{3} = E_n(1, \cos(\phi_{n3}+\phi_{H}) , \sin(\phi_{n3}+\phi_{H}), 0)\,, \notag \\
&p_{4}= E_n(1, \cos(\phi_{n4}+\phi_{H}), \sin(\phi_{n4}+\phi_{H}), 0)\,, \notag \\
&\vdots \notag \\
&p_{n}= E_n(1, \cos(\phi_{nn}+\phi_{H}), \sin(\phi_{nn}+\phi_{H}), 0)\,, \notag \\
&p_s = k_s(1, \cos \phi_s , \sin \phi_s, 0)\,,
\end{alignat}
with $E_n = 2E/(n-2)$ and $\phi_{nm} = \pi(2m-3)/(n-2)$. The momenta $p_1$ to 
$p_n$ can then be used directly to evaluate the $n$-parton amplitude. For the 
computation of the $(n+1)$-parton process we assume the recoil of the emitted 
soft-gluon to be absorbed by the dipole spanned by partons $3$ and 
$4$. The momenta of partons $3$ and $4$ that enter the $(n+1)$-parton 
amplitude are then given by
\begin{alignat}{5}
& p'_3 = p_3 - p_s +  \frac{p_3 \cdot p_s}{ p_4 \cdot( p_3 - p_s)}  p_4 \,,
\notag \\
& p'_4 = \left( 1 -\frac{p_3 \cdot p_s}{ p_4 \cdot( p_3 - p_s)} \right) p_4  \, .
\end{alignat}

We define with $R_s$ the inverse ratio between an $n+1$-parton matrix element 
squared and its ``sum-over-dipoles'' approximation
\begin{alignat}{5}
R_s = & \frac{\as}{\pi} \trace \left[ H_n \sum_{i<j} \T_i  \cdot \T_j
	\frac{p_i \cdot p_j}{ p_i\cdot p_s \, p_j \cdot p_s} \right] 
	\frac{1}{\trace \; \left( c\, H_{n+1}  \right) }. \label{eq:Rs_def}
\end{alignat}
The QCD coupling $\as$ is assumed fixed here. Factorization of QCD matrix elements 
implies that in the limit of soft-gluon kinematics, i.e., 
$\lambda_s = k_s / ( 2E ) \to 0$, we have
\begin{equation}\label{eq:Rs_limit}
 \lim_{\lambda_s\to 0} R_s =1. 
\end{equation}
This result is in fact independent of the underlying Born kinematics, and in 
particular independent of the angle $\phi_H$ through which we rotate our 
hard-parton configuration, Eq.~\eqref{eq:hard_soft_fac_test}.  Depending 
on $\phi_H$, the value of $R_s$ for finite $\lambda_s$ may be larger or 
smaller than one.  Taking the limit in \eq{eq:Rs_limit}, we provide a 
strong consistency check on the elements of $\mathbf{\Gamma}$ and 
$H_n$ for the $n$-parton process as well as $c$ and $H_{n+1}$ for 
the $n+1$ parton configuration. This applies to elements which 
have a non-vanishing hard contribution.

In order to expose this property of the full matrix element, 
we sample over $\phi_H$ in discrete steps assuring that the momentum 
$p_s$ does not get collinear to any other parton. This is sufficiently 
satisfied by requiring that $\phi_H$ is not an integer multiple of 
$\phi_s$. In practice we take $\phi_s = \pi / 7$, and sample 
$\phi_H = N\pi/10$ over $N=0,\dots,9$.  

\FIGURE[t]{
  \centering\centerline{
    \includegraphics[scale=.58]{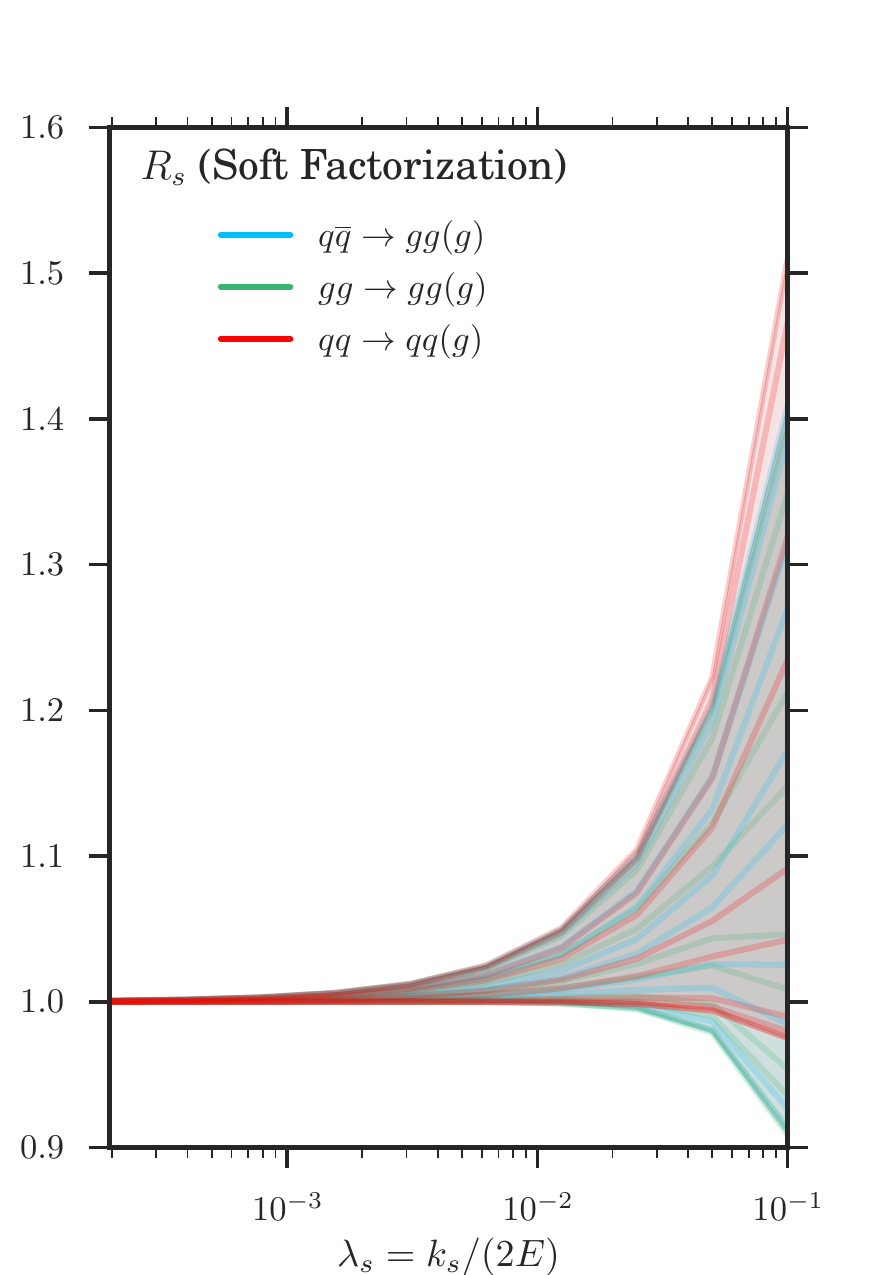}
    \hspace{-0.4cm}
    \includegraphics[scale=.58]{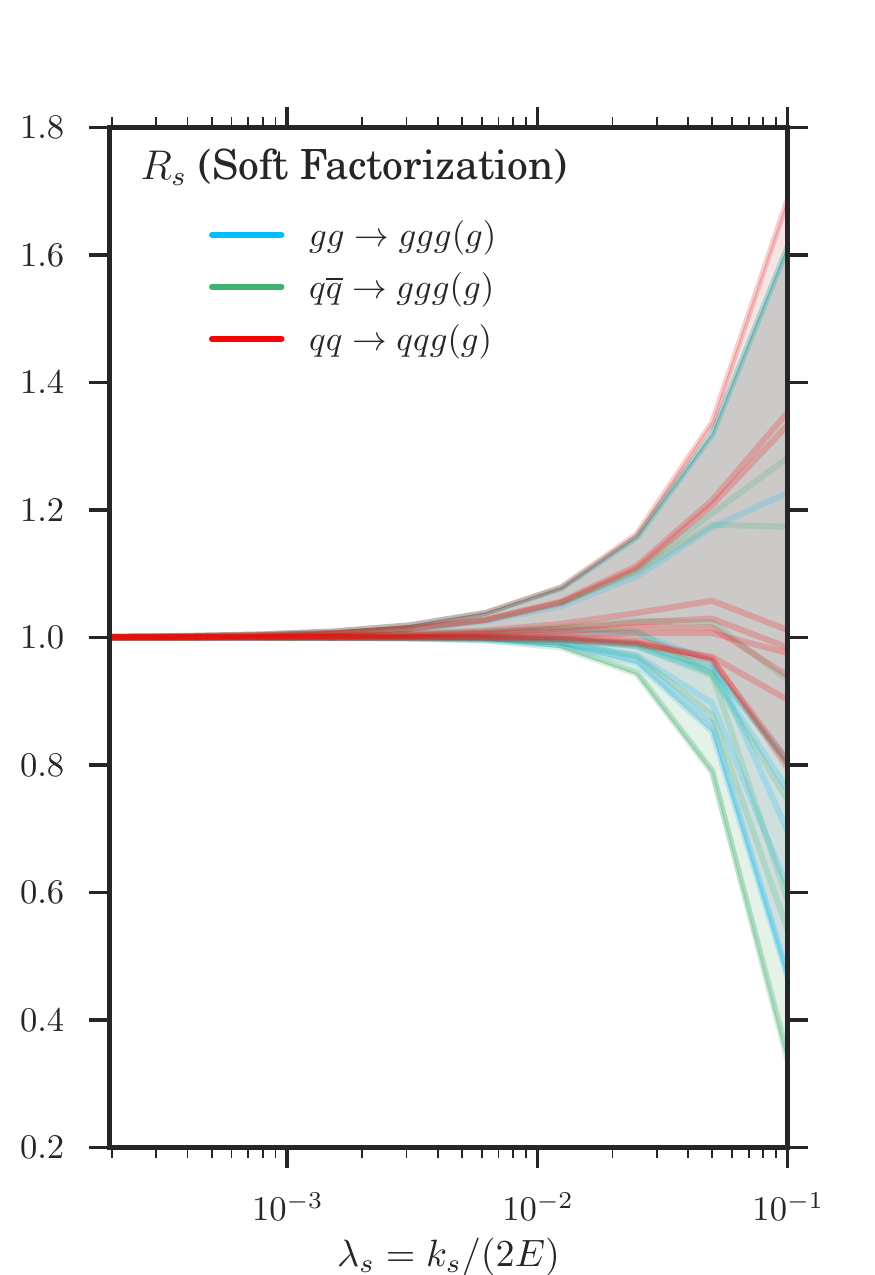}
    \hspace{-0.4cm}
    \includegraphics[scale=.58]{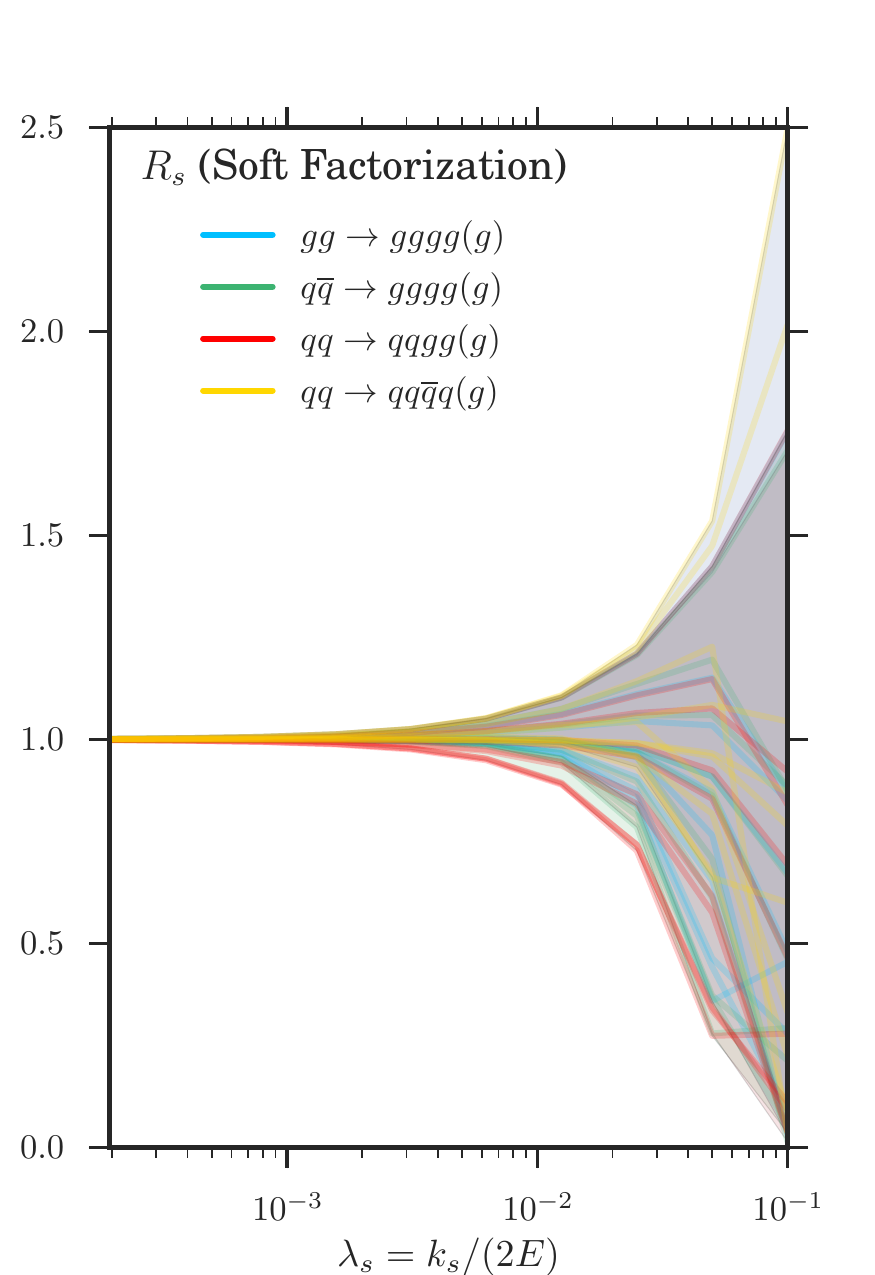}
  \caption{\label{fig:SC4567}%
(Left) Ratio of the sum-over-dipole dressed $4$-parton approximation 
to the exact $5$-parton matrix elements for different partonic 
subprocesses, cf. Eq.~(\ref{eq:Rs_limit}). (Middle) $R_s$ ratio 
for sum-over-dipole dressed $5$-parton over full $6$-parton 
configurations. (Right) Same but for $6/7$-parton matrix elements. 
}} 	
}

In \fig{fig:SC4567} we display the results of our checks for soft-gluon 
emission off $4$- (left), $5$-parton (middle) and $6$-parton (right) amplitudes. 
The last case provides a non-trivial check also on the $7$-parton color metric 
and hard matrix, entering through the denominator of Eq.~\eq{eq:Rs_limit}. 
For completeness we collect in App.~\ref{app:mp_bases} the properties of 
the color bases used for the various processes. By rotating the respective 
Born kinematics on the 
circle in the transverse plane, we can verify that the individual 
coefficients of each dipole are exactly matched in the full matrix element 
as $\lambda_s \to 0$. For a specific phase-space point these coefficients 
could be individually very small thus not providing a sufficient test. 
The results show a strong dependence on the underlying kinematic 
configuration in addition to the considered parton flavors. However, for
sufficiently small $\lambda_s$ in all cases $R_s$ approaches unity, proving
correctness of the ingredients for the soft function $\mathcal{S}$.

\subsection{Soft evolution of multi-parton squared amplitudes}

Having proven correctness of our color-metric evaluation and the
corresponding hard-matrix decomposition, we shall now study the 
full soft function ${\mathcal{S}}(\xi)$ given in Eq.~\eqref{soften}
for multi-parton processes. In particular we probe the dependence 
on the evolution variable $\xi$ and compare to the limiting case 
of $\nc\to \infty$, that closely resembles the approximation used 
in parton-shower simulations.   While the $\nc = \infty$ anomalous dimension
is computed explicitly in the trace basis, it amounts to only 
a non-vanishing contribution to $\T_i \cdot \T_j$ for basis 
elements 
which have partons $i$ and $j$ color adjacent.

We begin by computing $\mathcal{S}(\xi)$ for several multiplicities at
benchmark kinematics, that lie on a circle in the transverse
plane at $z=0$. For the $2\to n$ processes we parameterize the 
momenta as 
\begin{alignat}{5}
p_1 &= E(1,0,0,1)\notag\,, \\
p_2 &= E(1,0,0,-1) \notag\,, \\
p_3 &= E_n(1,\cos\phi_{n3},\sin\phi_{n3},0)\notag\,, \\
p_4 &= E_n(1,\cos\phi_{n4},\sin\phi_{n4},0) \notag\,, \\
&\vdots \notag \\
p_n &= E_n(1,\cos\phi_{nn},\sin\phi_{nn},0)\,,
\end{alignat}
where again $E_n = 2E/(n-2)$ and $\phi_{nm} = \pi(2m-3)/(n-2)$. The 
soft function $\mathcal{S}$ depends on the kinematics merely through 
ratios of momentum invariants (cusp angles), 
such that when considering fixed $\as$ the direct dependence on $E_n$ 
vanishes. 

\FIGURE{
  \centering\centerline{
  \includegraphics[scale=.58]{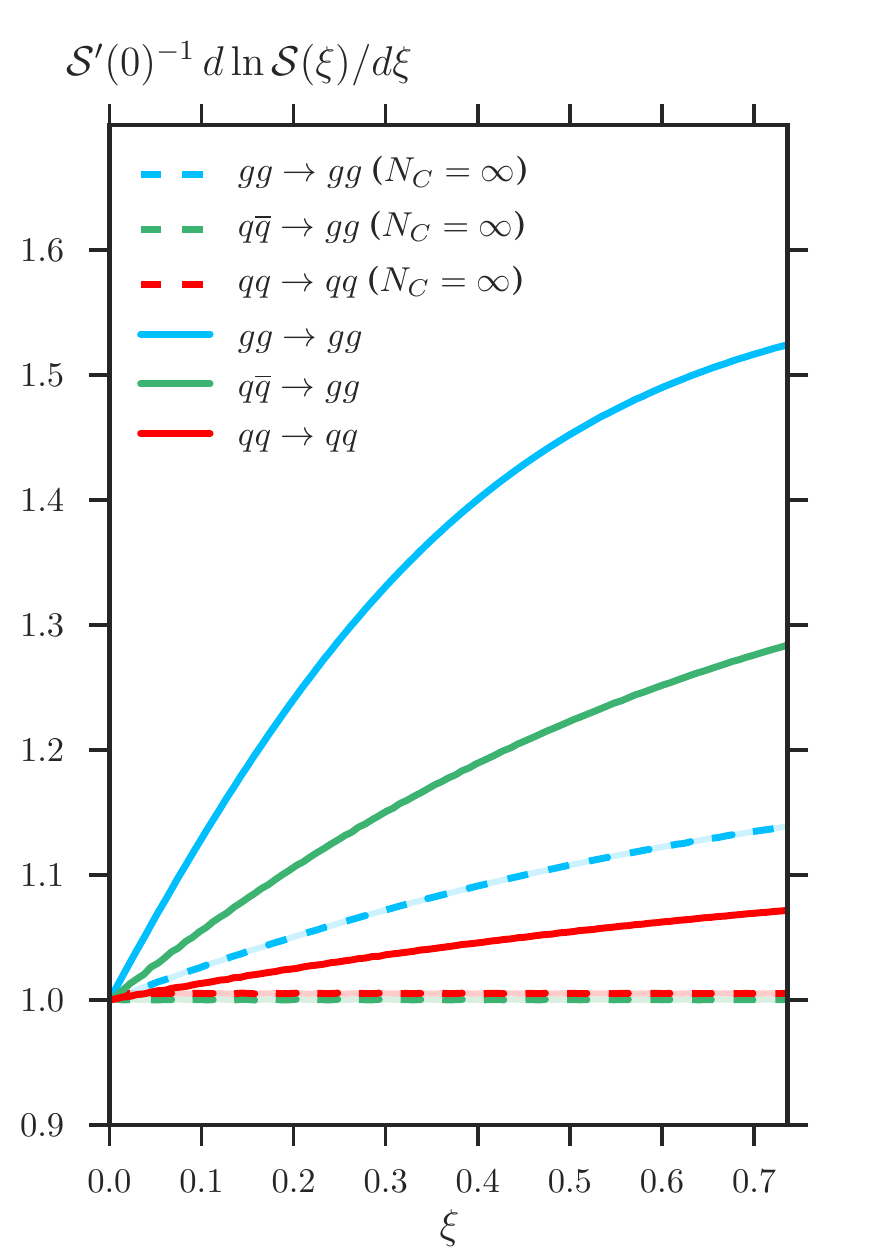}
  \hspace{.4cm}
  \includegraphics[scale=.58]{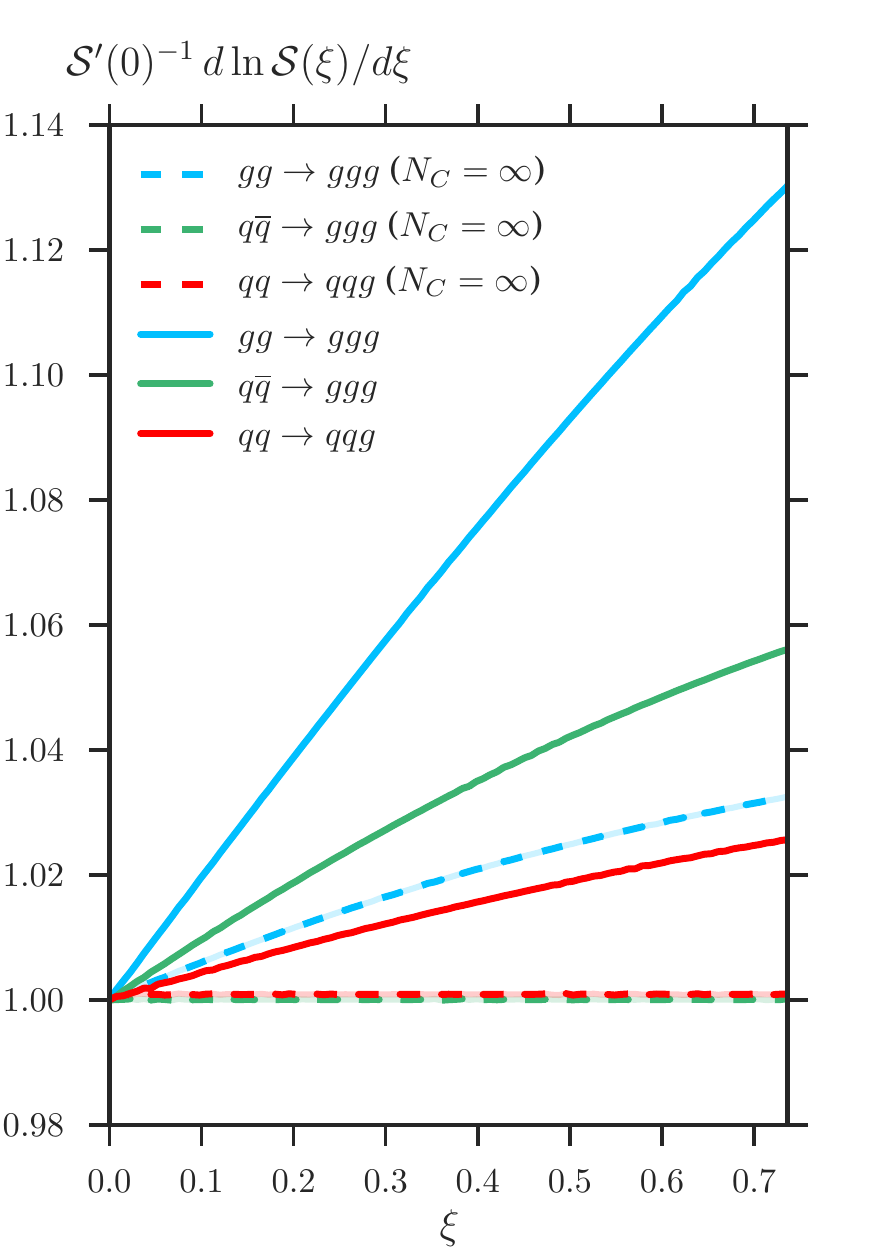} 
  \caption{\label{fig:S45}%
  Dependence of the soft function $\mathcal{S}$ on the evolution variable $\xi$
  for $2 \to 2$ (left) and $2 \to 3$ (right) parton configurations. For all
  processes parton momenta on a circle in the transverse plane at $z=0$ are 
  considered.}}
}
\FIGURE{
  \centering\centerline{
  \includegraphics[scale=0.58]{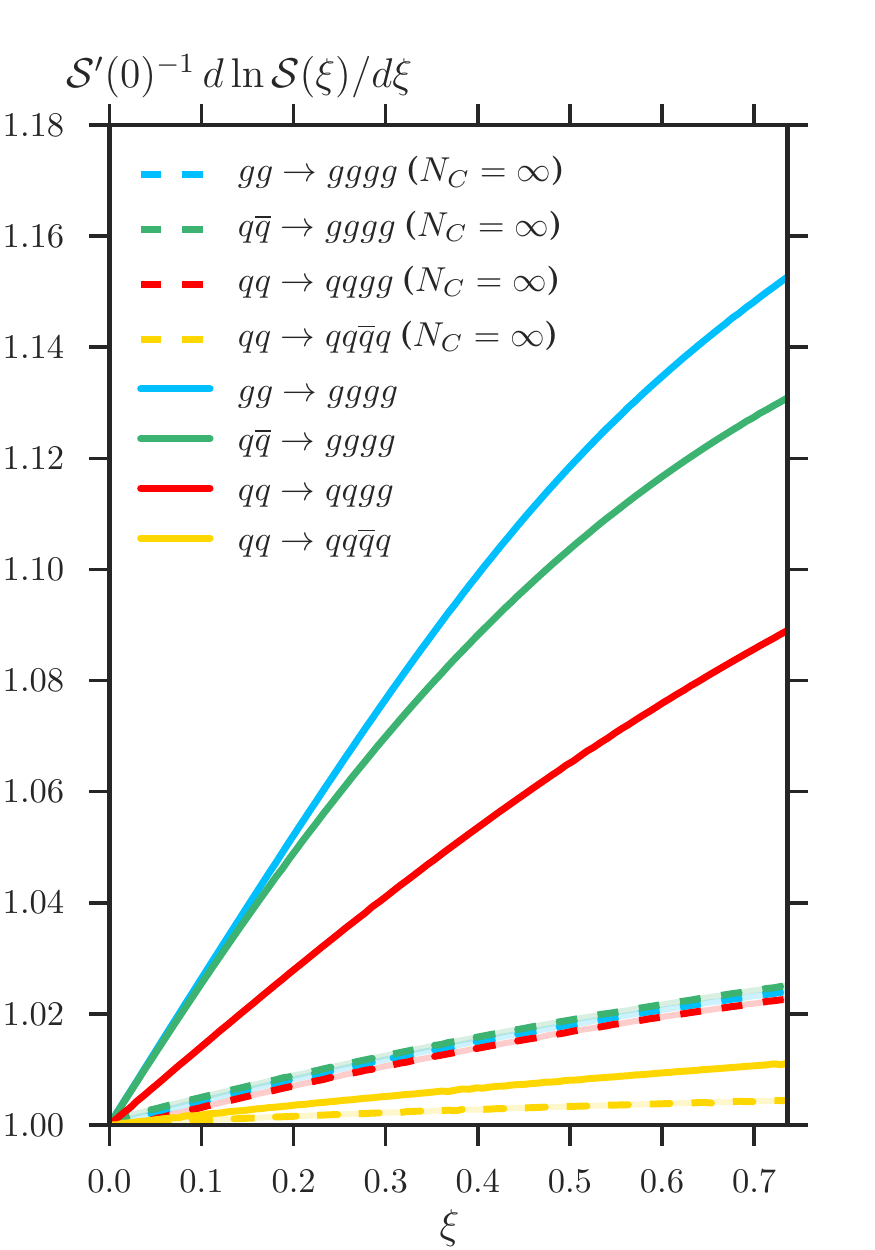}
  \hspace{.4cm}
  \includegraphics[scale=0.58]{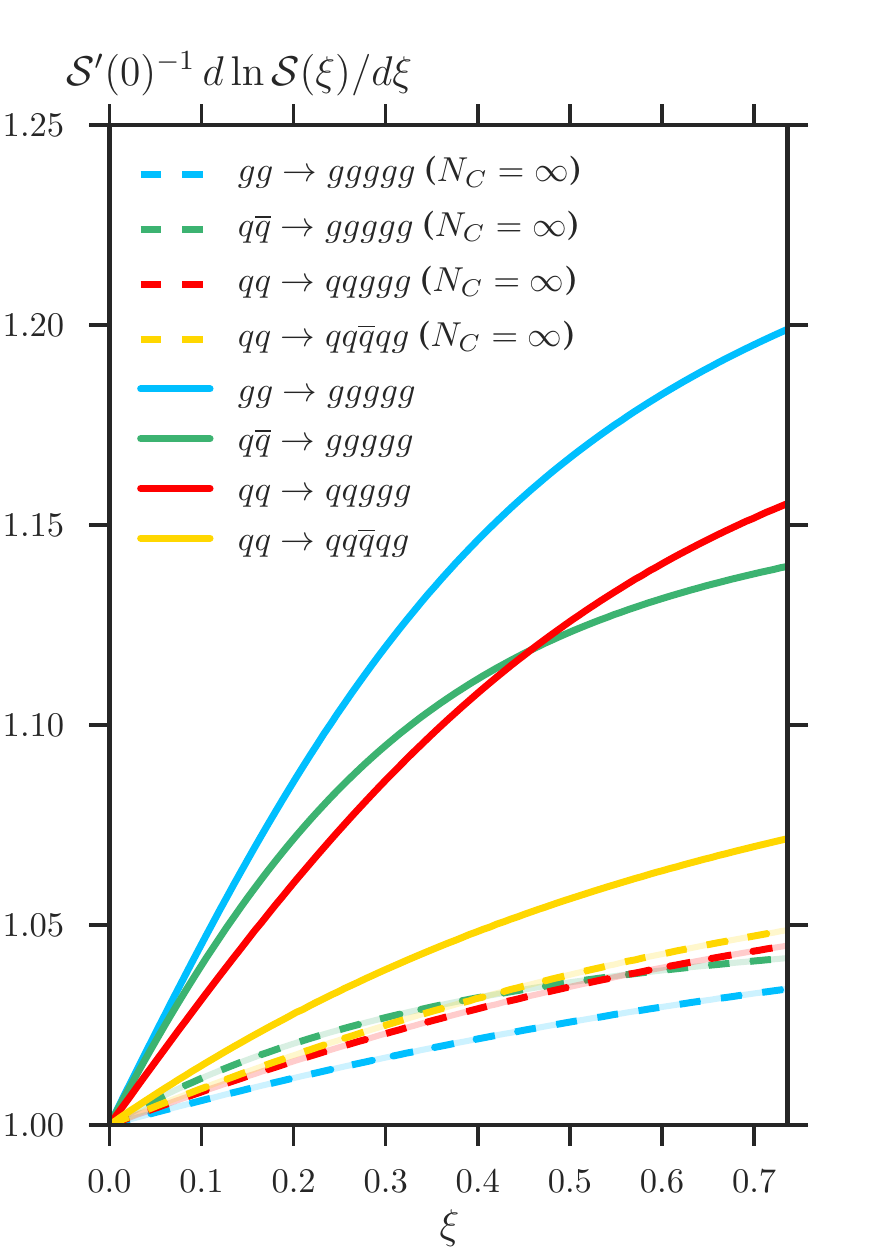}
  \caption{\label{fig:S67}%
    Dependence of the soft function $\mathcal{S}$ on the evolution variable 
    $\xi$ for $2 \to 4$ (left) and $2 \to 5$ (right) parton configurations. 
    For all processes parton momenta on a circle in the transverse plane at 
    $z=0$ are considered.}} 	
}

In \fig{fig:S45} and \fig{fig:S67} we present results for the
$\xi$-dependence of the soft function for various parton channels, 
both for $\nc=3$ (solid curves) and for the limit $\nc\to \infty$ 
(dashed curves). Depicted is the variation of $\ln {\mathcal{S}}(\xi)$
with $\xi$, where we scaled each curve such that it intersects 
with the ordinate at one.  
We observe a non-trivial 
$\xi$ dependence for all processes when considering the full-color 
treatment. For the given phase-space configurations the full result
shows a stronger variation with $\xi$ than the large-$\nc$ estimate.
This originates from taking into account all off-diagonal elements 
in the soft anomalous dimension. In particular for processes involving 
gluons the limit $\nc \to \infty$ approximates the full result poorly.  

Let us discuss the general behaviour of our results 
in the large-$\nc$ case. In this case, all non-diagonal entries of the soft anomalous dimension vanish and we can simply write
\begin{equation} \label{largeNcparam}
\frac{1}{\mathcal{S}'(0)} \frac{d \log\mathcal{S}(\xi)}{d\xi}
=
\left(\dfrac{\sum_{i=1}^{n} h_{ii}} 
{\sum_{i=1}^{n} \lambda_i \, h_{ii}} \right)
\dfrac{\sum_{i=1}^{n} h_{ii} \lambda_i \exp(\lambda_i\xi)} 
{ \sum_{i=1}^{n} h_{ii} \exp(\lambda_i \xi) } \qquad (\text{at large\,-$\nc$})
\end{equation}
for an $n$-dimensional color space where $\lambda_i$ and $h_{ii}$ are the diagonal entries of 
$\Gamma_{\alpha\beta}$ and $H^{\alpha\beta}$ respectively. 
Both $\lambda_i$ and $h_{ii}$ are positive because they correspond to non-interfering squared amplitudes. Consequently, Eq.~(\ref{largeNcparam}) is a monotonically increasing function $\xi$, for all underlying Born configurations. However, at finite $\nc$, the full matrix structure persists and the behaviour is not neccesarily monotonic due to off-diagonal 
interfering contributions.

\FIGURE[t!]{
  \centering\centerline{
  \includegraphics[scale=0.58]{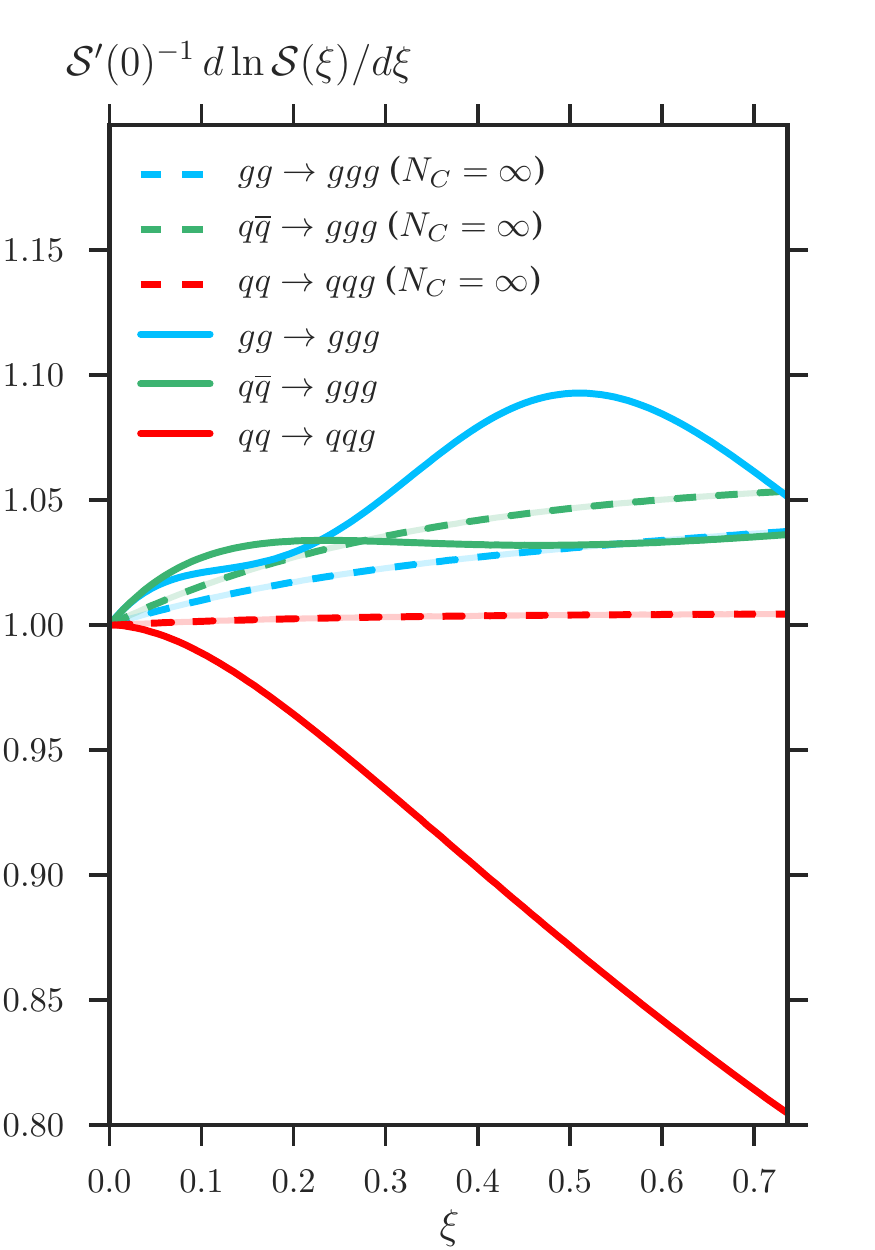}
  \hspace{-.4cm}
  \includegraphics[scale=0.58]{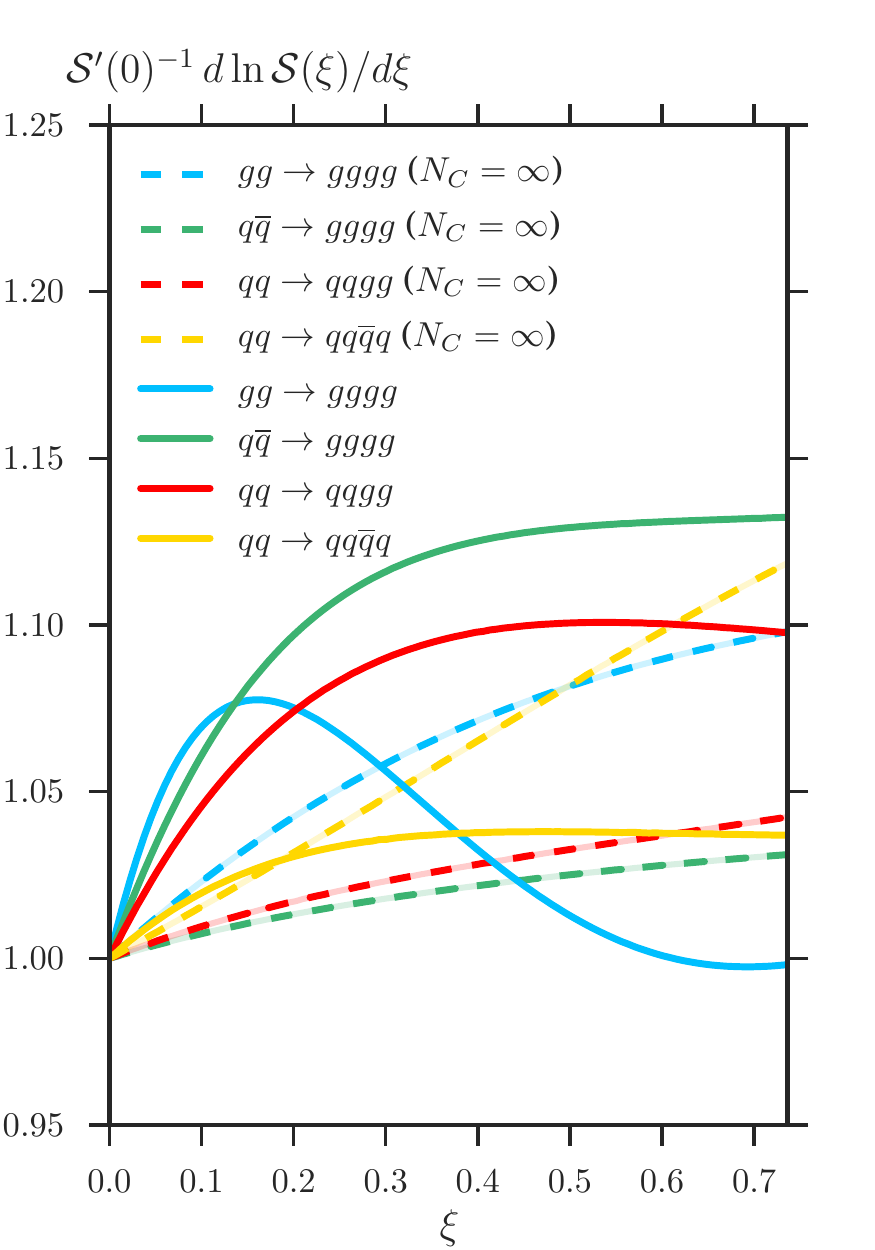}
  \hspace{-.4cm}
  \includegraphics[scale=0.58]{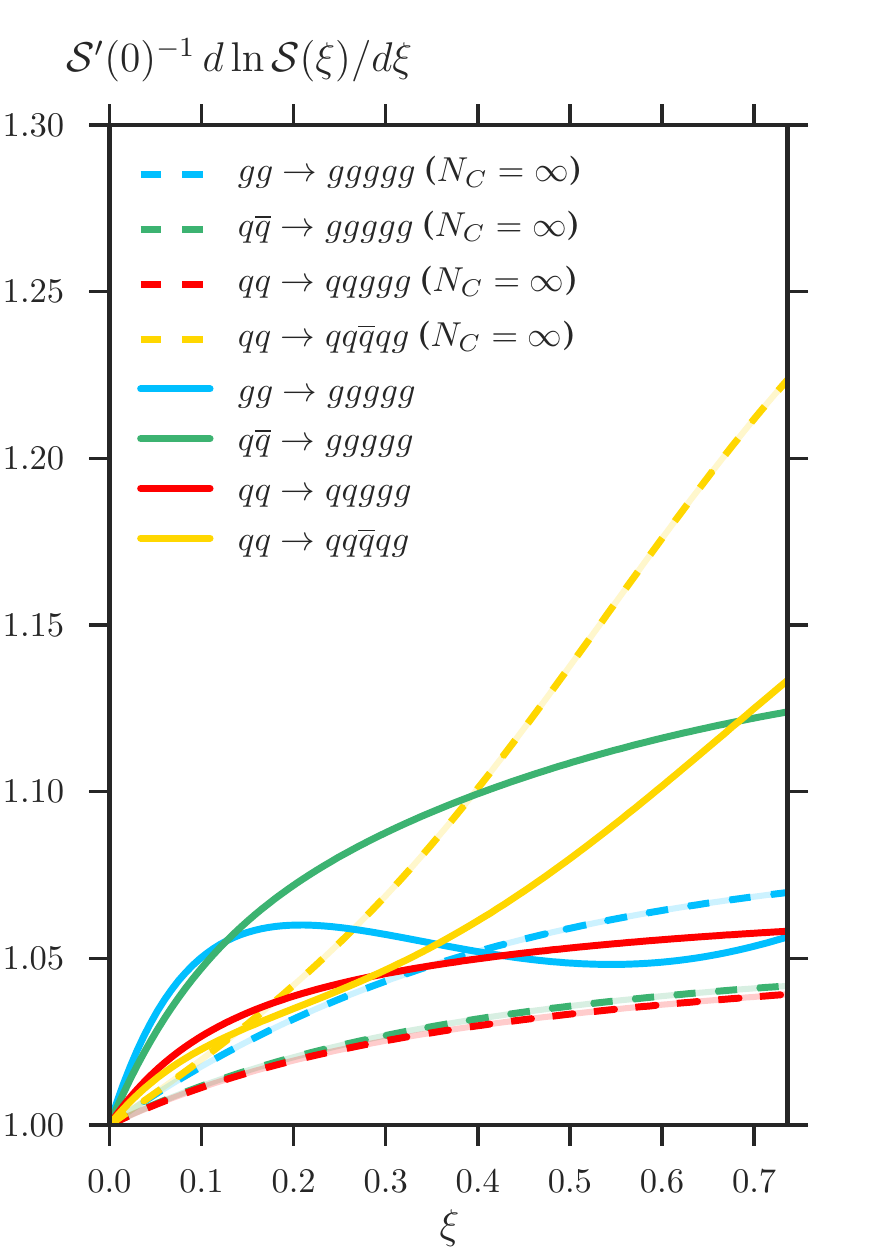}
  \caption{\label{fig:SNC1}%
    Soft function for $2\to 3$ (left), $2\to 4$ (middle) and $2\to 5$ (right)
    parton configurations, for kinematics with all final-state momenta in
    the plane of the beam. 
}} 	
}

To also check kinematic configurations with particles at non-zero rapidity,
we considered the above kinematics but rotated by an angle $\pi/2$ about
the $y$-axis. This results in momenta that span a circle in the $y-z$ plane
at $x=0$. The corresponding results can be found in \fig{fig:SNC1}. Again, 
while the behaviour for $\nc \to \infty$ are necessarily monotonically increasing 
functions of $\xi$ this is not true for finite $\nc$, due to non-vanishing 
interference effects of different color flows. Accordingly, the large-$\nc$
approximation can in general also result in an overestimate of the soft 
function. To properly account for the highly non-trivial dependence on 
the parton kinematics and the evolution variable the soft function needs
to be evaluated with its full color dependence, i.e. $\nc=3$. However, 
to fully quantify the importance of finite-$\nc$ effects not just 
the soft-function contribution but the full physical observable needs 
to be considered.

%% file: inputs/resum.tex
§\section{Towards phenomenology}\label{sec:resummation}
In the first part of this paper, we have presented a new method to deal with the soft evolution of processes with many colored legs that provides a high degree of automation. Moreover, we have realized an implementation of this method that uses color-partial amplitudes extracted from the matrix-element generator \comix and evolves them according to the soft anomalous dimension Eq.~(\ref{eq:gamma-ind}), thus obtaining an efficient way of evaluating the soft function given in Eq.~(\ref{eq:logS-basis-ind}).

The aim of this second part is to create a framework in which the soft function $\mathcal{S}$ in Eq.~(\ref{eq:logS-basis-ind}) can be used for phenomenological studies. Let us generically call $v$ an observable that measures the ``distance'' from the lowest-order kinematics. In the context of jet studies, $v$ can be thought of as an observable describing internal jet properties, e.g.\ masses,  angularities, energy correlation functions) or as an observable measuring the radiation outside the leading jets, e.g.\ event shapes or inter-jet radiation. When $v$ is small, logarithms $L=\ln \frac{1}{v}$ are large and
resummation becomes a more efficient organization of the perturbative expansion than fixed-order perturbation theory. Furthermore, we have to consistently match the two approaches to obtain reliable predictions for the entire range of the observable $v$: 
\beq \label{eq:matching}
\frac{d \sigma^\text{matched}}{d v}=\frac{d \sigma^\text{resummed}}{d v}+\left(\frac{d \sigma^\text{fixed-order}}{d v}-\frac{d \sigma^\text{expanded}}{d v}\right).
\eeq
The first term in the expression above is computed to some logarithmic accuracy, typically next-to-leading log (NLL) but not infrequently to NNLL, while the second one is computed at a given order in the strong coupling (state of the art is typically NLO). The last term represents the expansion of the resummed distribution to NLO and avoids double counting. The last two terms are affected by large logarithms and are in fact separately divergent in the limit $v\to0$. However, their combination yields a finite remainder, called the matching term.  Although conceptually trivial, computing the matching term is often numerically inefficient because it involves the separate evaluation of fixed-order contribution and expanded resummation in regions of phase space corresponding to soft and/or collinear emissions. It would be preferable to generate the finite remainder directly. 

\subsection{Resummed distributions} \label{sec:resum_with_caesar}
Resummed calculations are usually performed for the so-called cumulative distribution, i.e.\ the integral of the differential distribution up to a certain value $v$ of the observable under consideration:
\begin{align} \label{eq:fact}
\frac{d \Sigma(v)}{d \mathcal{B}} &=\frac{1}{\sigma}\int_0^v \frac{d^2 \sigma}{d \mathcal{B}  d v'} d v'\nonumber  \\  &=\sum_{\begin{subarray}{c}
      \text{partonic}\\ \text{configurations}\\ \delta
      \end{subarray}}
       \frac{d \sigma_0^{(\delta)}}{d \mathcal{B} } e^{Lg_1^{(\delta)}(\alpha_s L)+g_2^{(\delta,\mathcal{B})}(\alpha_s L) +\dots} \left[1+\O{} \right],          
\end{align}
where $d \mathcal{B} $ indicates that the expression above is fully differential in the Born kinematics.
Here we focus our attention on the NLL approximation of $\ln \Sigma$, i.e.\ we consider the functions $g_1^{(\delta)}$ and $g_2^{(\delta,\mathcal{B})}$ in Eq.~(\ref{eq:fact}), while dropping the non-logarithmic term in square brackets. 
The inclusion of this constant contribution is necessary in order to achieve what often is referred to as NLL$^\prime$ accuracy. We note that to this logarithmic accuracy such contribution, although flavor-sensitive, can be averaged over the different color flows. Furthermore, we note that this constant term can be extracted from NLO calculations as implemented, for instance, with the \powheg method~\cite{Nason:2004rx,*Frixione:2007vw}, which has been automated in the \sherpa framework in Ref.~\cite{Hoche:2010pf}~\footnote{We acknowledge discussions with Gavin Salam, Mrinal Dasgupta and Emanuele Re over this point.}.
For our discussion, we follow the formalism developed in the context of the program \caesar~\cite{Banfi:2001bz,*Banfi:2003je,*Banfi:2004nk,*Banfi:2004yd}, which allows one to resum global event shapes in a semi-automated way. With a couple of generalizations, the \caesar framework is sufficient for our purposes. Furthermore, we will also briefly discuss some differences in the structure of the resummation that arise when dealing with non-global observables~\cite{Dasgupta:2001sh,*Dasgupta:2002bw} at the end of this section.

We consider processes which at Born level feature $n$ hard massless partons (legs) and $m$ color singlets (e.g.\ photons, Higgs or electroweak bosons) and we denote the set of Born momenta with $\{p\}$. Following Refs.~\cite{Banfi:2001bz,*Banfi:2003je,*Banfi:2004nk,*Banfi:2004yd} we consider positive-definite observables $V$ that measure the difference in the energy-momentum flow of an event with respect to the Born configuration, where $V(\{p\})=0$. For a single emission with momentum $k$, which is soft and collinear to leg $l$, the observable $V$ is parametrized as follows~\footnote{In principle we should consider the set of momenta $\{\tilde{p}\}$ after recoil, but this effect is beyond the NLL accuracy aimed for here~\cite{Banfi:2001bz,*Banfi:2003je,*Banfi:2004nk,*Banfi:2004yd}.}
\begin{equation}\label{caesar-obs}
V\left(\{\tilde{p}\};k\right)= d_l \left(\frac{k_t^{(l)}}{Q} \right)^{a}e^{-b_l \eta^{(l)}}g_l\left(\phi^{(l)} \right),
\end{equation}
where $k_t^{(l)}$, $\eta^{(l)}$ and $\phi^{(l)}$ denote transverse momentum, rapidity and azimuth of the emission, all measured with respect to parton $l$. $Q$ is the hard scale of the process which we set equal to the partonic centre of mass energy, i.e.\ $Q^2=s$.
It is then possible to write the resummed exponent in Eq.~(\ref{eq:fact}) in terms of the coefficients $a$, $b_l$, $d_l$ and $g_l(\phi)$ that specify the behavior of the observable in the presence of a soft and collinear emission. 

In particular, while the LL function $g_1^{(\delta)}$ is diagonal in color, one of the contributions that enter the NLL function $g_2^{(\delta,\mathcal{B})}$ is precisely the soft function, which, as discussed in Sec.~\ref{sec:resum},  has a matrix structure in color space, with complexity that increases with the number of hard partons.
Explicit formulae are collected in App.~\ref{app:caesar}.
The results of Sec.~\ref{sec:resum} provide an automated way of computing these contributions, thus extending the applicability of the \caesar framework to processes with an (in principle) arbitrary large number of hard partonic legs.

We conclude this discussion with a few remarks on non-global observables~\cite{Dasgupta:2001sh,*Dasgupta:2002bw}. 
Non-global logarithms arise for those observables that have sharp geometrical boundaries in phase space. They originate in wide-angle soft gluons that lie outside the region where the observable is measured, re-emitting softer radiation back into that region. The \caesar framework presented above is not sufficient to deal with this case and new ingredients need to be introduced. Most noticeably, the NLL function $g_2^{(\delta,\mathcal{B})}$ receives a new contribution coming from correlated gluon emission~\footnote{We should mention that the particular choice of the algorithm used to define jets can influence the resummation structure at the level of $g_2^{(\delta,\mathcal{B})}$. This discussion refers to a jet algorithm, like for instance anti-$k_t$~\cite{Cacciari:2008gp}, which in the soft limit behaves as a rigid cone.}.
Because of their soft and large-angle nature, non-global logarithms have a complicated color structure. However, for phenomenological purposes, their resummation can be performed in the large-$N_C$ limit~\cite{Dasgupta:2001sh,*Dasgupta:2002bw,Banfi:2002hw,Schwartz:2014wha}, thus trivializing the color structure again. 
Recent studies suggest a way of performing this resummation at finite $N_C$~\cite{Hatta:2013iba}. We believe that the methodology for performing all-order calculations with many hard legs can also prove useful in the application of those methods to LHC phenomenology. However, we leave this investigation for future work.
Finally, we point out that while the color structure of the soft anomalous dimension $\mathbf\Gamma$ for non-global observables is formally the same as in Eq.~(\ref{eq:gamma-ind}), the coefficients of the $\T_i\cdot \T_j$ are observable-dependent, because of non-trivial limits for the azimuth and rapidity integrals.

\subsection{Automated matching}\label{sec:automated-matching}
In order to avoid double counting when matching a resummed calculation to a fixed-order one, we need to consider the expansion of the resummation. In this paper, we are concerned with matching to tree-level matrix elements, thus we have to consider the expansion of the NLL resummed distribution to $\mathcal{O}\left(\as \right)$
\begin{equation}\label{expansionLO}
\frac{d}{d L}\frac{d \Sigma^{(\delta)}}{d \mathcal{B}}=  \frac{2 \as}{\pi} \frac{d}{d L } \left[\frac{G_{12}}{2} L^2+ G_{11} L \right] + \mathcal{O}\left(\as^2 \right) ,
\end{equation}
with $\as=\as(\mu_R^2)$ and $L=\ln\left(1/ v \right)$. 

If the resummation is performed within the \caesar formalism, which is summarized for convenience in App.~\ref{app:caesar}, one is able to expressed the coefficients $G_{12}$ and $G_{11}$ in terms of the coefficients that parametrize the observable in Eq.~(\ref{caesar-obs}). An explicit calculation leads to
\begin{align}\label{expansionLO_coeff}
G_{12}=&- \sum_{l=1}^n \frac{C_l}{a (a+b_l)} \nonumber\\ 
G_{11}=&- \left[\sum_{l=1}^n C_l \left(\frac{B_l}{a+b_l}+\frac{1}{a(a+b_l)}\left(\ln \bar{d}_l-b_l \ln \frac{2 E_l}{Q} \right)+\frac{1}{a} \ln\frac{Q_{12}}{Q} \right) \right. \nonumber\\ &+\left.
\frac{1}{a}\frac{{\rm Re} [ \Gamma_{\alpha \beta}]\, H^{\alpha \beta}}{c_{\alpha \beta}H^{\alpha \beta}}
+\sum_{l=1}^{n_\text{initial}} \frac{\int_{x_l}^1 \frac{dz}{z} P_{l k}^{(0)}\left( \frac{x_l}{z}\right) q^{(k)}(z,\mu_{F}^2) }{2(a+b_l) q^{(l)}(x_l,\mu_{F}^2)}
 \right]\,.
\end{align}
%Expanding the resummed prediction in the \caesar formalism to $\mathcal{O}\left(\as \right)$ one obtains (see App.~\ref{app:caesar} for details) 
%\begin{align}\label{expansionLO}
%\frac{d}{d L}\frac{d \Sigma^{(\delta)}}{d \mathcal{B}}&=- \frac{2 \as}{\pi} L \sum_{l=1}^n \frac{C_l}{a (a+b_l)} \nonumber\\ &-\frac{2 \as}{\pi} \left[\sum_{l=1}^n C_l \left(\frac{B_l}{a+b_l}+\frac{1}{a(a+b_l)}\left(\ln \bar{d}_l-b_l \ln \frac{2 E_l}{Q} \right)+\frac{1}{a} \ln\frac{Q_{12}}{Q} \right) \right. \nonumber\\ &+ \left.
%\frac{1}{a}\frac{{\rm Re} [ \Gamma_{\alpha \beta}]\, H^{\alpha \beta}}{c_{\alpha \beta}H^{\alpha \beta}}
%+\sum_{l=1}^{n_\text{initial}} \frac{\int_{x_l}^1 \frac{dz}{z} P_{l k}^{(0)}\left( \frac{x_l}{z}\right) q^{(k)}(z,\mu_{F}^2) }{2(a+b_l) q^{(l)}(x_l,\mu_{F}^2)}
% \right],
%\end{align}
%with $\as=\as(\mu_R^2)$ and $L=\ln\left(1/ v \right)$.
Our aim is to compute $G_{12}$  and the first term in $G_{11}$ by integrating collinear splitting functions
in a Monte-Carlo approach over suitably defined regions of phase space. This procedure is similar to
next-to-leading order subtraction techniques. It allows to combine the matching terms with real-emission
matrix elements point-by-point in the real-emission phase space, and provides therefore a quasi-local 
cancellation of large logarithms in the matching\footnote{The cancellation is not necessarily local
  because the parametrization of the observable in terms of kinematical variables may differ from 
  the actual real-emission kinematics.}. We use an existing implementation of the Catani--Seymour 
dipole-subtraction method in \sherpa{}~\cite{Gleisberg:2007md} as the basis for our implementation.
The remaining terms proportional to $C_l$ in $G_{11}$ are generated by using the color-correlated 
Born amplitudes only and multiplying with the analytic expression for $\log\bar{d}_l-b_l \ln(2 E_l/Q)$
(or $\log(Q_{12}/Q)$) and the relevant prefactors. The generation of the second line in 
Eq.~\eqref{expansionLO_coeff} is described in detail below.

The dipole-subtraction method of Ref.~\cite{Catani:1996jh,*Catani:1996vz} is based on the soft and collinear 
factorization properties of tree-level matrix elements. In the collinear limit we can write
\begin{equation}\label{eq:cs_collfac}
  \begin{split}
    &|\mathcal{M}_0(1,\ldots,i,\ldots,j,\ldots,n)|^2\;\overset{i,j\to\text{collinear}}{\longrightarrow}\\
    &\quad\frac{8\pi\mu^{2\varepsilon}\alpha_s}{2p_ip_j}\,\langle m_0(1,\ldots,ij,\ldots,n)|\,
    \hat{P}_{ij,i}(z,k_T,\varepsilon)\,|m_0(1,\ldots,ij,\ldots,n)\rangle\;.
  \end{split}
\end{equation}
The splitting operators $\hat{P}_{ij,i}$ describe the branching $ij\to i,j$ as a function of 
the light-cone momentum fraction $z=n p_i/n (p_i+p_j)$, with $n$ an auxiliary vector, 
and the transverse momentum $k_T^2=2p_ip_j\,z(1-z)$. The splitting operators depend non-trivially 
on the helicity of the combined parton, $ij$, but they have a trivial color structure.
In the soft limit, the matrix element factorizes as
\begin{equation}\label{eq:cs_softfac}
  \begin{split}
    &|\mathcal{M}_0(1,\ldots,j,\ldots,n)|^2\;\overset{j\to\text{soft}}{\longrightarrow}\;
    -\sum_{i,k\neq i}\frac{8\pi\mu^{2\varepsilon}\alpha_s}{p_ip_j}\\
    &\quad\times\langle m_0(1,\ldots,i,\ldots,k,\ldots,n)|
    \frac{{\bf T}_i \cdot {\bf T}_k\; Q_{ik}}{Q_{ij}+Q_{kj}}\,|m_0(1,\ldots,i,\ldots,k,\ldots,n)\rangle\;.
  \end{split}
\end{equation}
The color insertion operators ${{\bf T}_i \cdot \bf T}_k$ are the same as in Eq.~\eqref{eq:gamma-ind}.
The full insertion operator has a trivial helicity dependence. 
Ref.~\cite{Catani:1996jh,*Catani:1996vz} combines the two above equations into a single 
factorization formula, which holds both in the soft and in the collinear region. 
The full matrix element is then approximated by a sum of dipole terms, which are defined as 
\begin{align}\label{eq:cs_fac}
    &\mathcal{D}_{ij,k}(1,\ldots,n)\,=\;-\frac{1}{2p_ip_j}\\
    &\quad\times\langle m_0(1,\ldots,ij,\ldots,k,\ldots,n)|
    \frac{{\bf T}_i \cdot{\bf T}_k}{{\bf T}_{ij}^2}\;\hat{V}_{ij,k}(z,k_T,\varepsilon)\,
      |m_0(1,\ldots,ij,\ldots,k,\ldots,n)\rangle\;.\nonumber
\end{align}
The insertion operators $\hat{V}_{ij,k}(z,k_T,\varepsilon)$ are based on the collinear splitting
operators $\hat{P}_{ij,i}(z,k_T,\varepsilon)$, and modified such that $Q_{ik}/(Q_{ij}+Q_{kj})$ 
is recovered in the soft limit. 

This formula is exploited for matching in the following way:
\begin{enumerate}
\item The color insertion operators are identical to the ones in the anomalous dimension ${\bf \Gamma}$.
Upon replacing $\hat{V}_{ij,k}(z,k_T,\varepsilon)$ by $2\log Q_{(ij)k}/Q_{12}$, and rescaling by $1/a$,
we obtain the term proportional to ${\rm Re} [ \Gamma_{\alpha \beta}]\, H^{\alpha \beta} / c_{\alpha \beta}H^{\alpha \beta}$
in Eq.~\eqref{expansionLO_coeff}. This is the only term with a non-trivial color structure.
\item The dipole splitting operators $\hat{V}_{ij,k}(z,k_T,\varepsilon)$ cancel the singularities
in the real-emission matrix element that we match to, in particular in the collinear limit, where
Eq.~\eqref{eq:cs_fac} reduces to Eq.~\eqref{eq:cs_collfac}. Upon replacing $\hat{V}_{ij,k}$ by
$\hat{P}_{ij,i}$, restricting doubly logarithmic terms to the appropriate region of phase space, and 
rescaling by $1/(a+b_l)$, we obtain $G_{12}$ and the term proportional to $B_l$ in $G_{11}$\footnote{
  For details on the definition of the integration region leading to $G_{11}$,
  see Ref.~\cite{Banfi:2001bz,*Banfi:2003je,*Banfi:2004nk,*Banfi:2004yd}.}.
\end{enumerate}
The factorization of the one-emission phase space is derived in Ref.~\cite{Catani:1996jh,*Catani:1996vz}
in terms of variables that represent scaled invariant masses and light-cone momentum fractions.
Based on these quantities we define two new variables, $v$ and $z$, as 
\begin{equation}
  \begin{aligned}
    v&=\left\{\begin{array}{cc}
    y_{ij,k} & \text{FF dipoles} \\[3mm]
    \displaystyle \frac{1-x_{ij,a}}{1-x_B} & \text{FI dipoles} \\[3mm]
    u_{i} & \text{IF dipoles} \\[3mm]
    \displaystyle \frac{v_{i}}{1-x_B} & \text{II dipoles}
    \end{array}\right.\;,\qquad
    &z&=\left\{\begin{array}{cc}
    \tilde{z}_j \text{ or } \tilde{z}_i & \text{FF dipoles} \\[2mm]
    \tilde{z}_j \text{ or } \tilde{z}_i & \text{FI dipoles} \\[3mm]
    \displaystyle \frac{1-x_{ik,a}}{1-x_B} & \text{IF dipoles} \\[4mm]
    \displaystyle \frac{1-x_{i,ab}}{1-x_B} & \text{II dipoles}
    \end{array}\right.\;.
  \end{aligned}
\end{equation}
In this context, $x_B$ is the Bj{\o}rken-$x$ of the Born process, pertaining to 
the initial-state leg for which the dipole is computed. The terms involving $x_B$ 
are included to obtain the correct integration range as compared to the resummation,
which is performed on Born kinematics, while Eq.~\eqref{eq:cs_fac} is computed for 
real-emission kinematics.

We restrict the phase space for the double-logarithmic term in soft-enhanced splitting
operators to the region $z^a>v$ (for terms singular as $z\to 0$). This corresponds 
to the requirement that the gluon rapidity in the rest frame of the radiating dipole 
be predominantly positive, and it generates the correct logarithmic dependence 
of $G_{11}$ in Eq.~\eqref{expansionLO_coeff}~\cite{Banfi:2001bz,*Banfi:2003je,*Banfi:2004nk,*Banfi:2004yd}.
More importantly it ensures that the soft-collinear singularity structure of the real-emission
matrix element is mapped out by the matching terms locally in the real-emission phase space.

Matching terms originating in dipoles with initial-state emitter or spectator 
are scaled by a ratio of parton densities, which accounts for the fact that the resummation 
starts from Born kinematics, while the dipole terms in Eq.~\eqref{eq:cs_fac} have real-emission kinematics.
This modification induces a single-logarithmic dependence on the observable, which is compensated 
by the explicit collinear counterterms in the expansion, i.e.\ the last term in Eq.~\eqref{expansionLO_coeff}. 
This term is computed independently.

\begin{figure}
  \centering
    \includegraphics[width=0.42\textwidth]{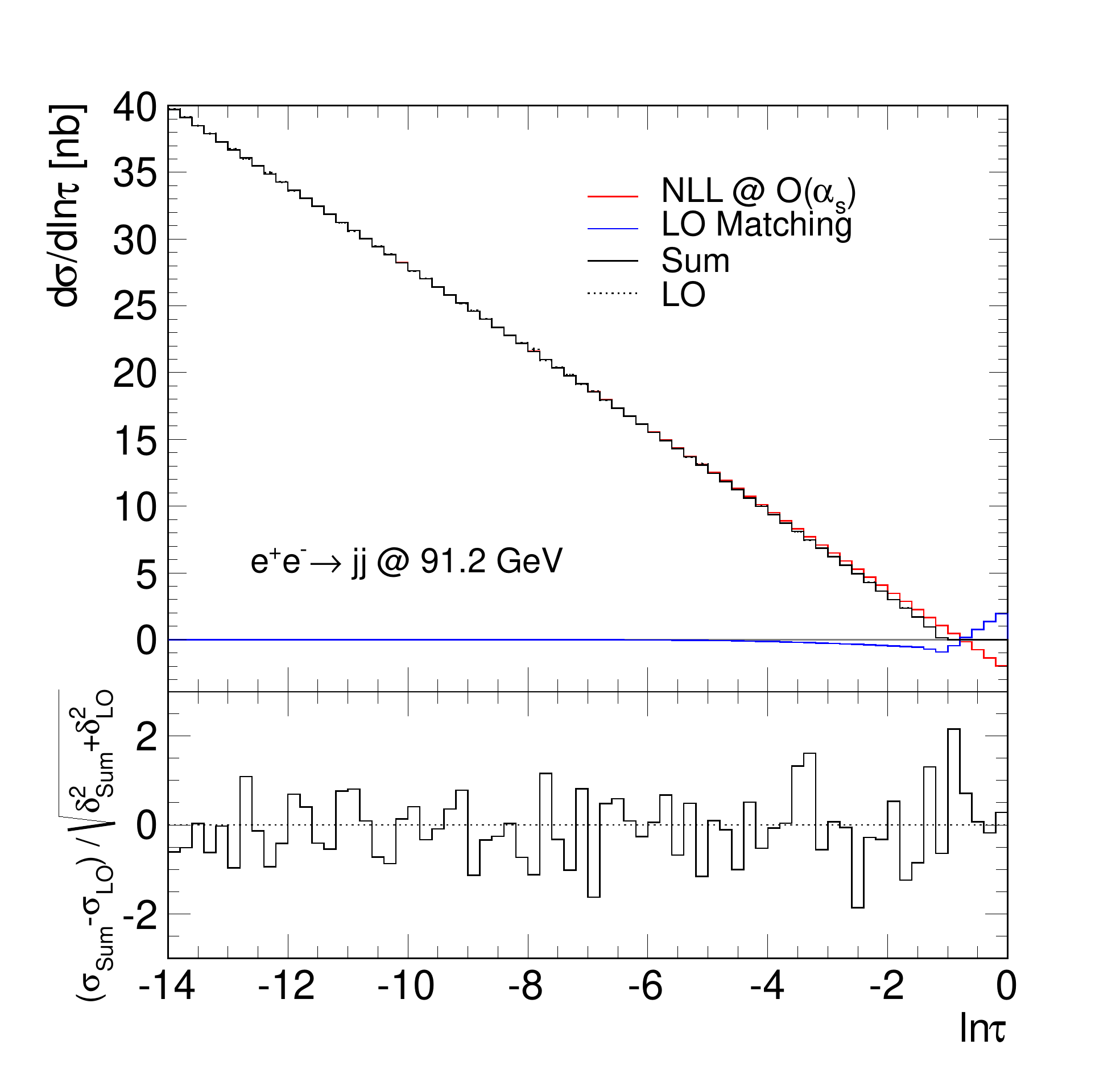}
    \includegraphics[width=0.42\textwidth]{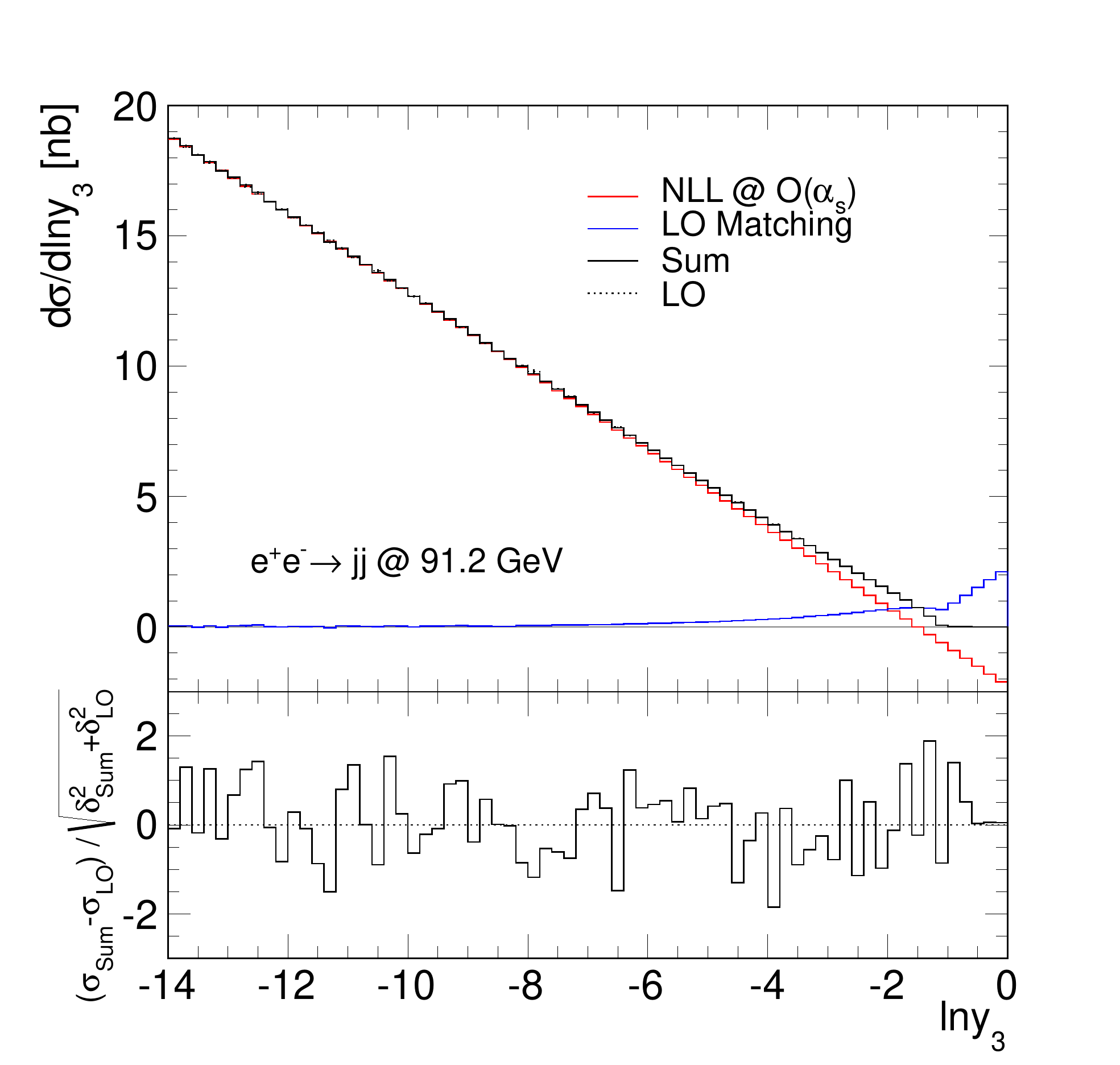}\\
    \includegraphics[width=0.42\textwidth]{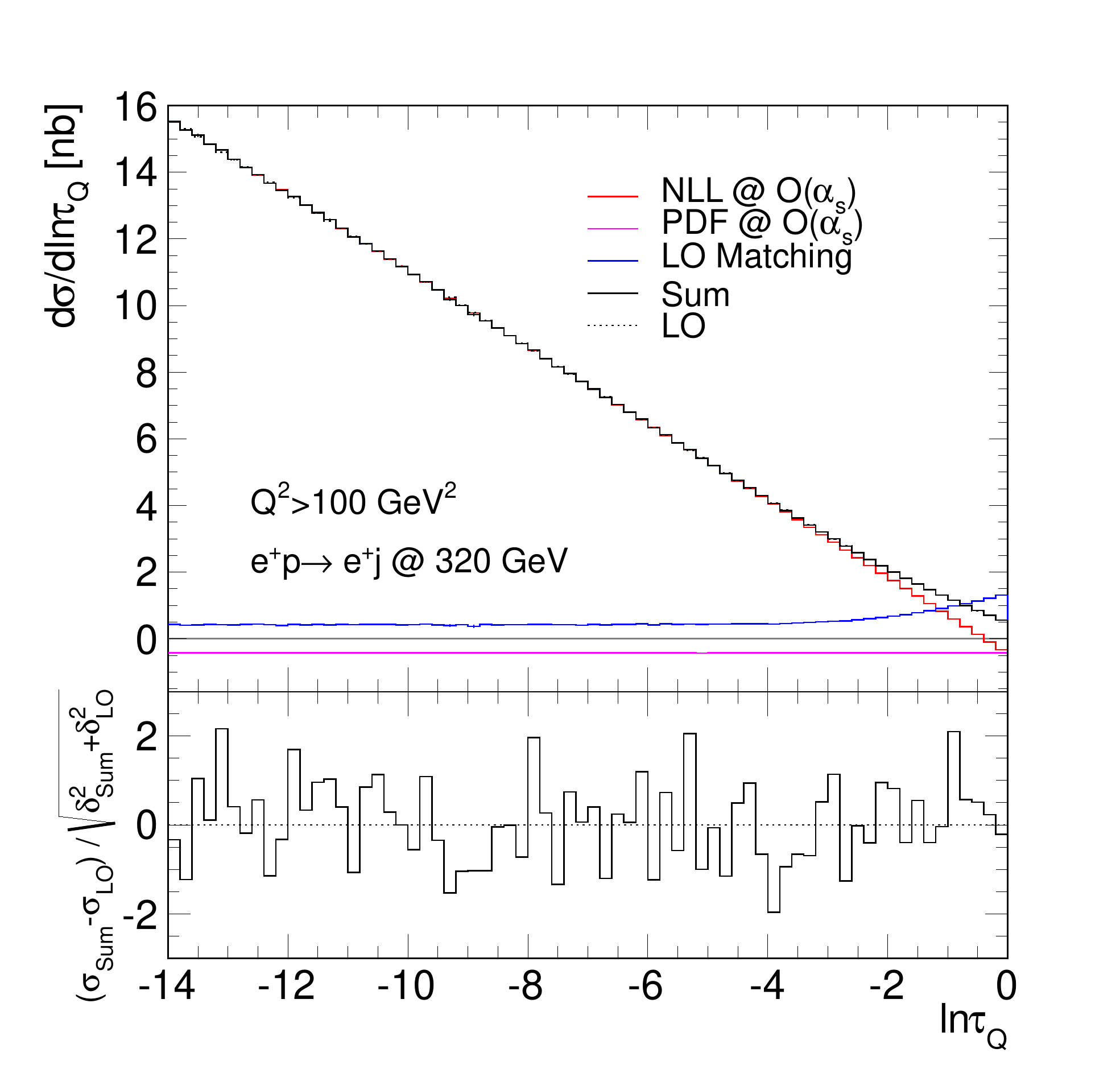}
    \includegraphics[width=0.42\textwidth]{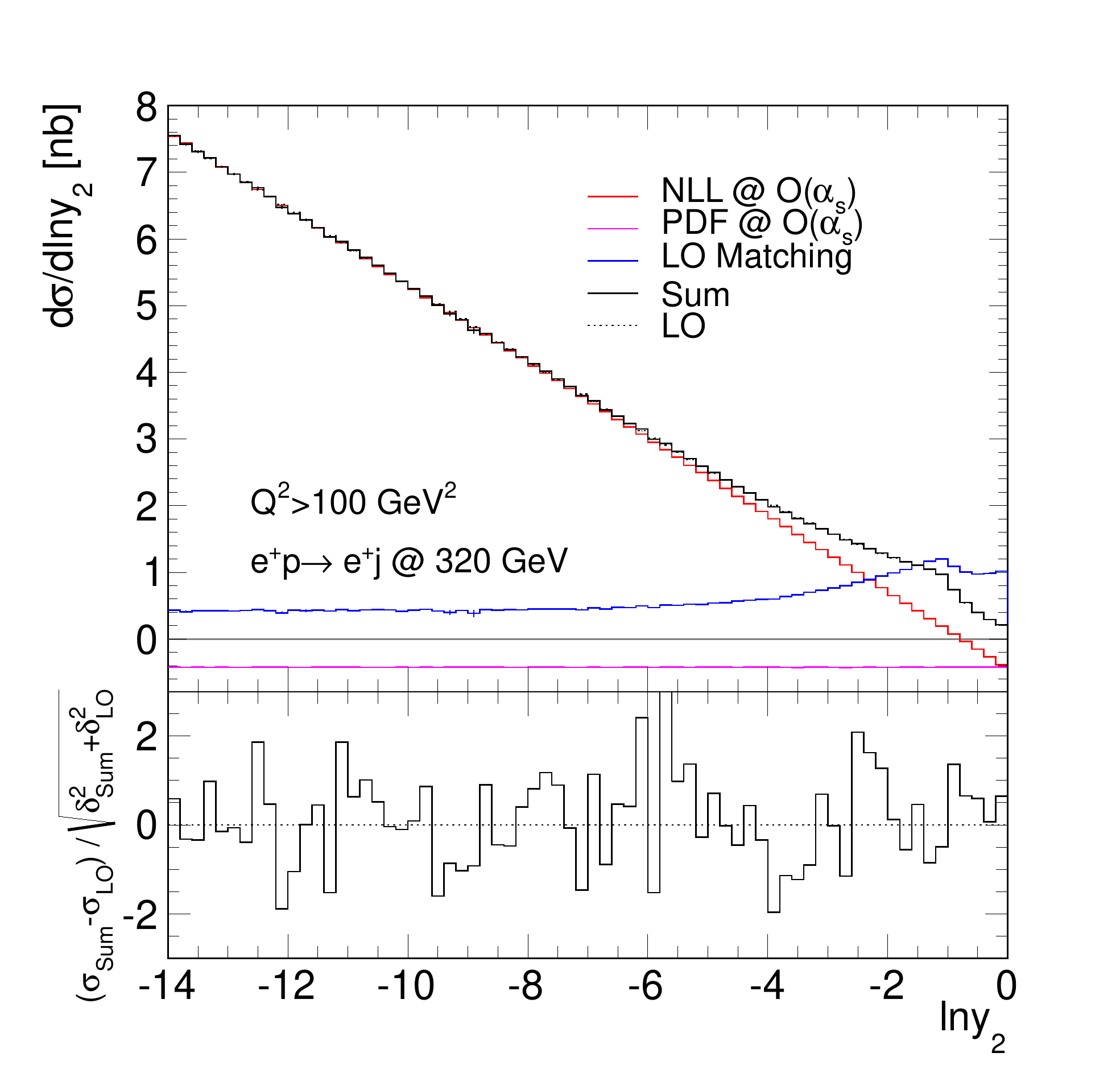}\\
    \includegraphics[width=0.42\textwidth]{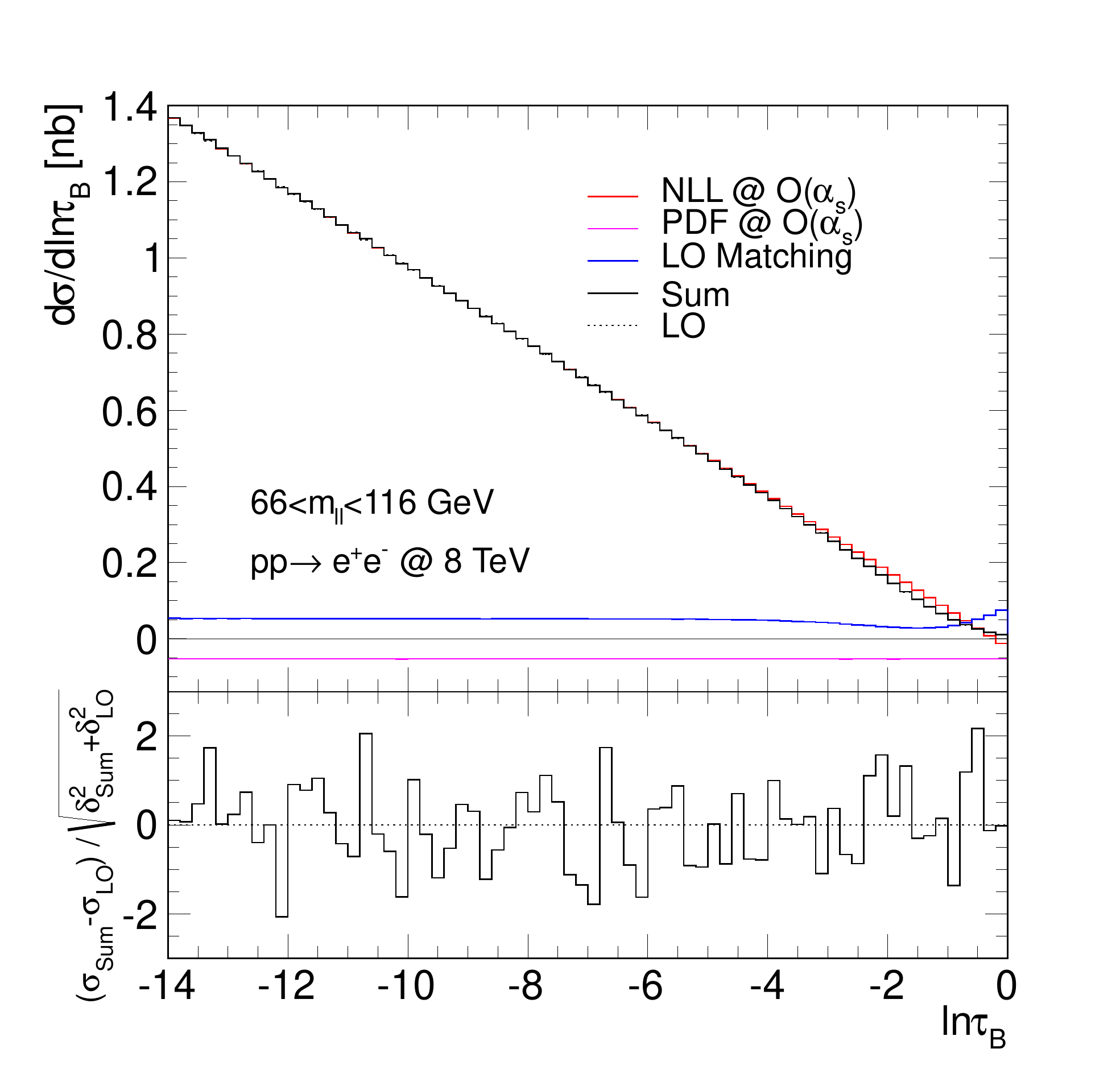}
    \includegraphics[width=0.42\textwidth]{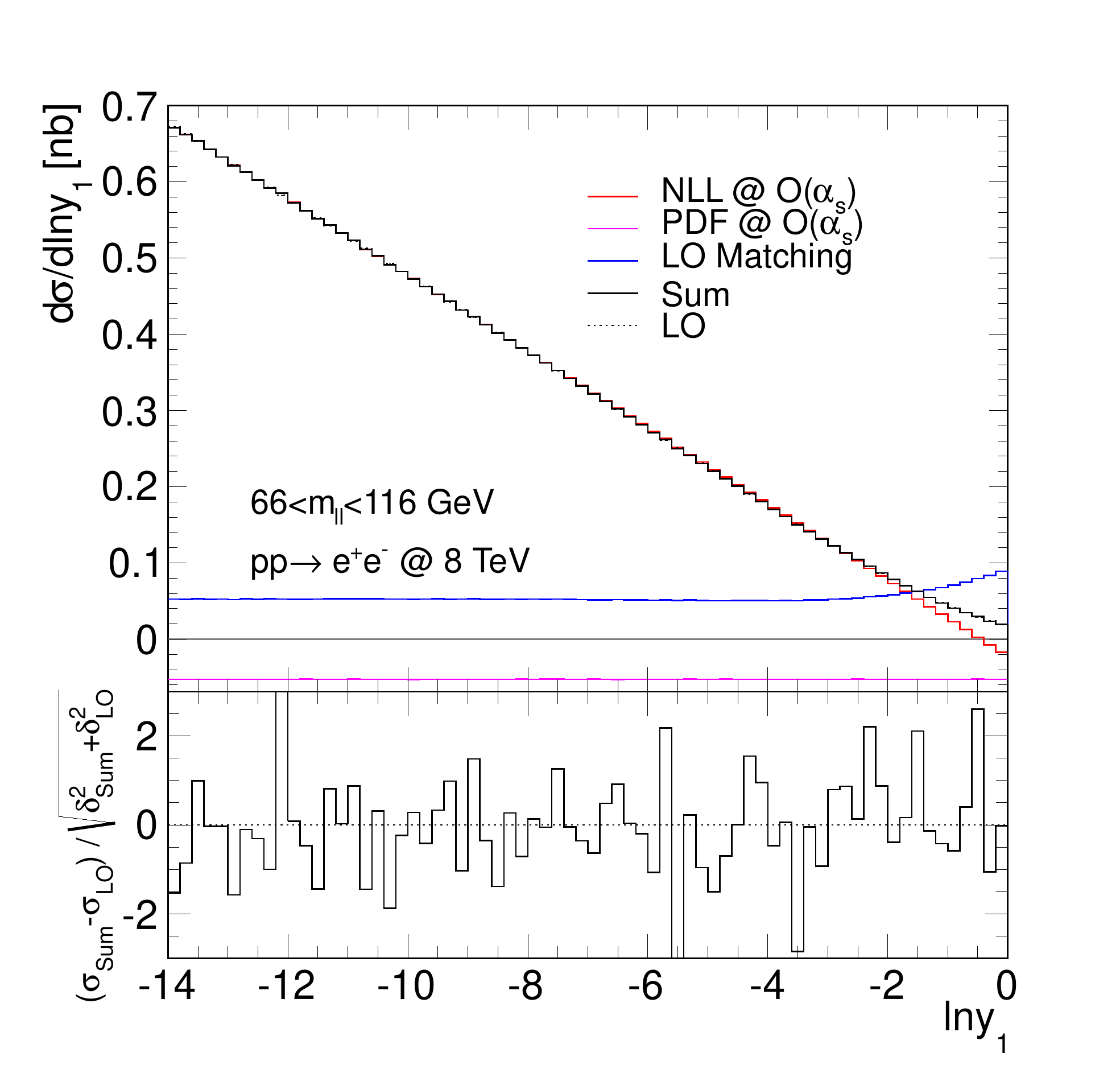}
  \caption{Test of the quasi-local matching procedure for hard processes with
    external quarks. Thrust (left panels) and leading jet rate (right panels)
    are compared between leading order and the first-order expanded 
    resummed and matched prediction for $e^+e^-\to q\bar{q}$ (top),
    $e^+ q\to e^+q$ (middle) and $q\bar{q}\to e^+e^-$ (bottom), all 
    mediated by photon and $Z$-boson exchange.
    \label{fig:matching_quarks}}
\end{figure}

\begin{figure}
  \centering
    \includegraphics[width=0.42\textwidth]{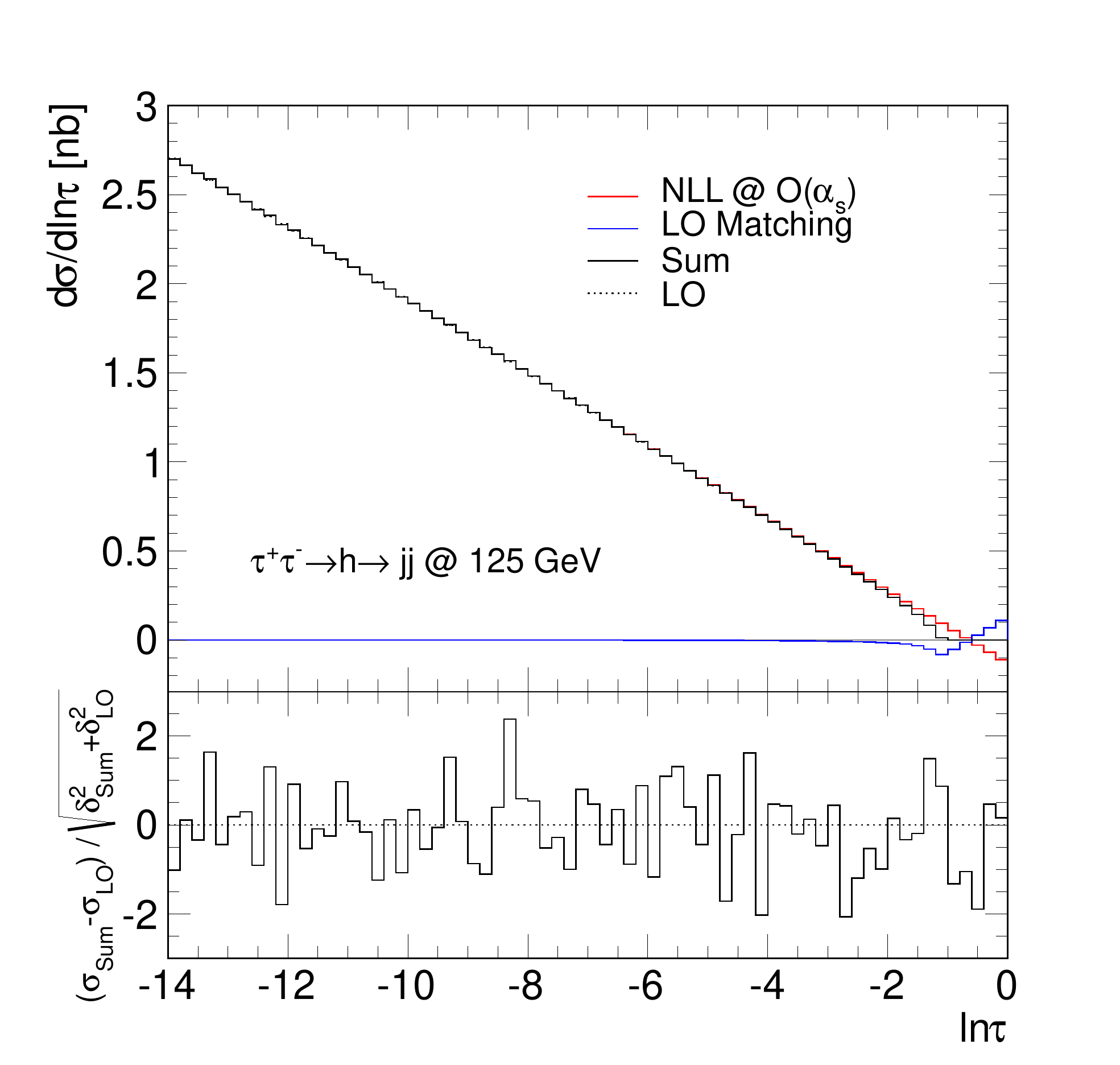}
    \includegraphics[width=0.42\textwidth]{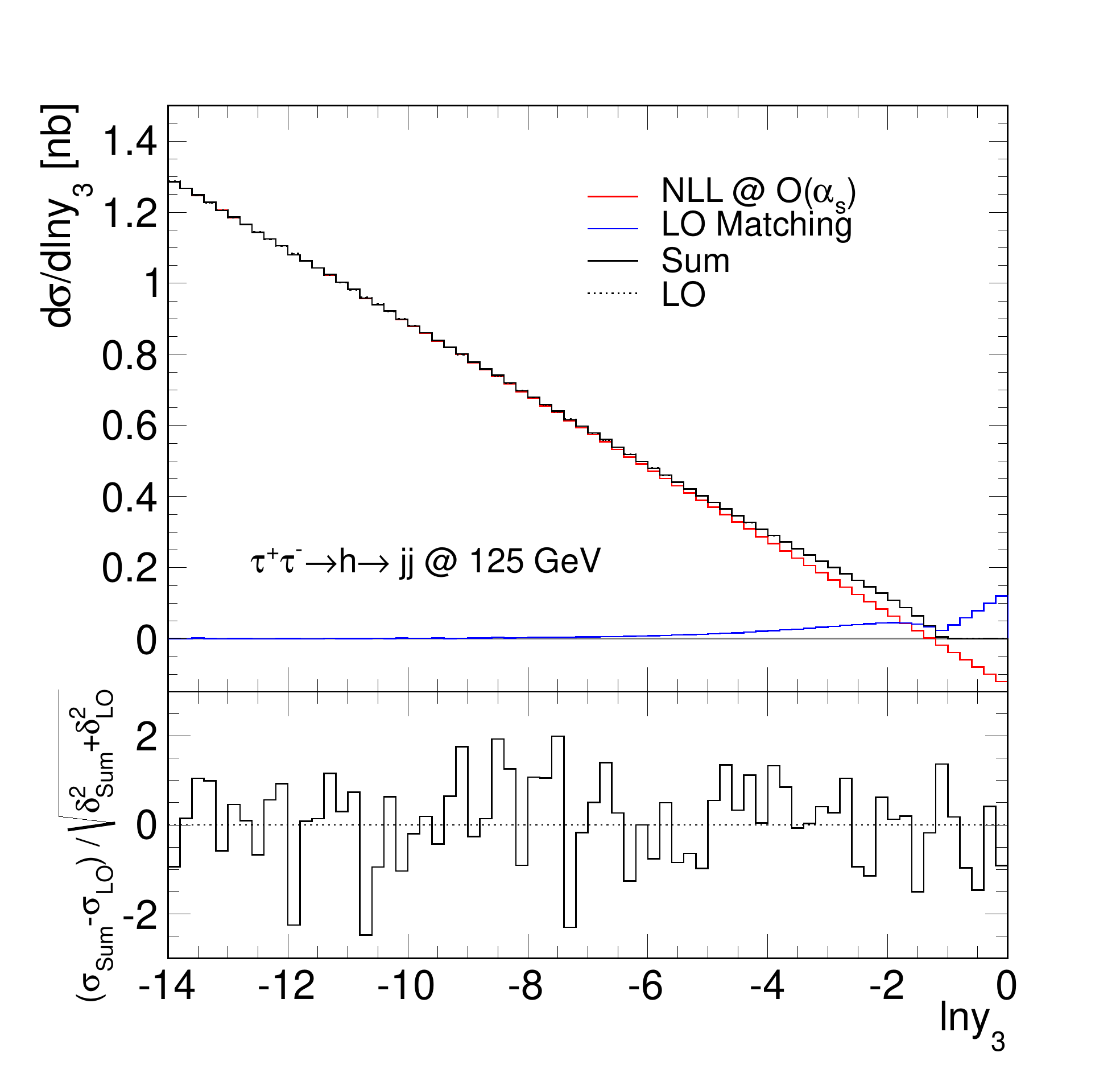}\\
    \includegraphics[width=0.42\textwidth]{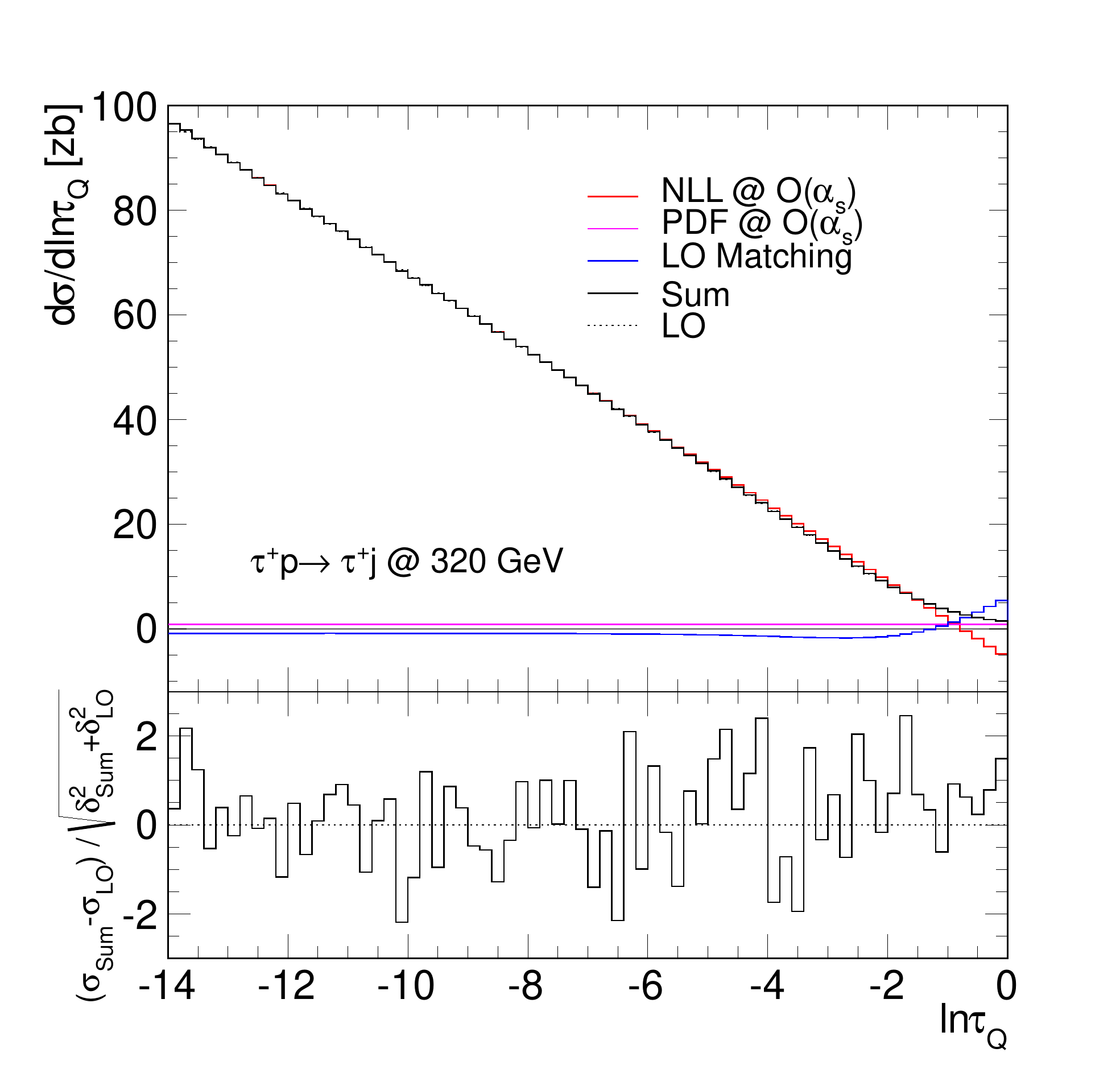}
    \includegraphics[width=0.42\textwidth]{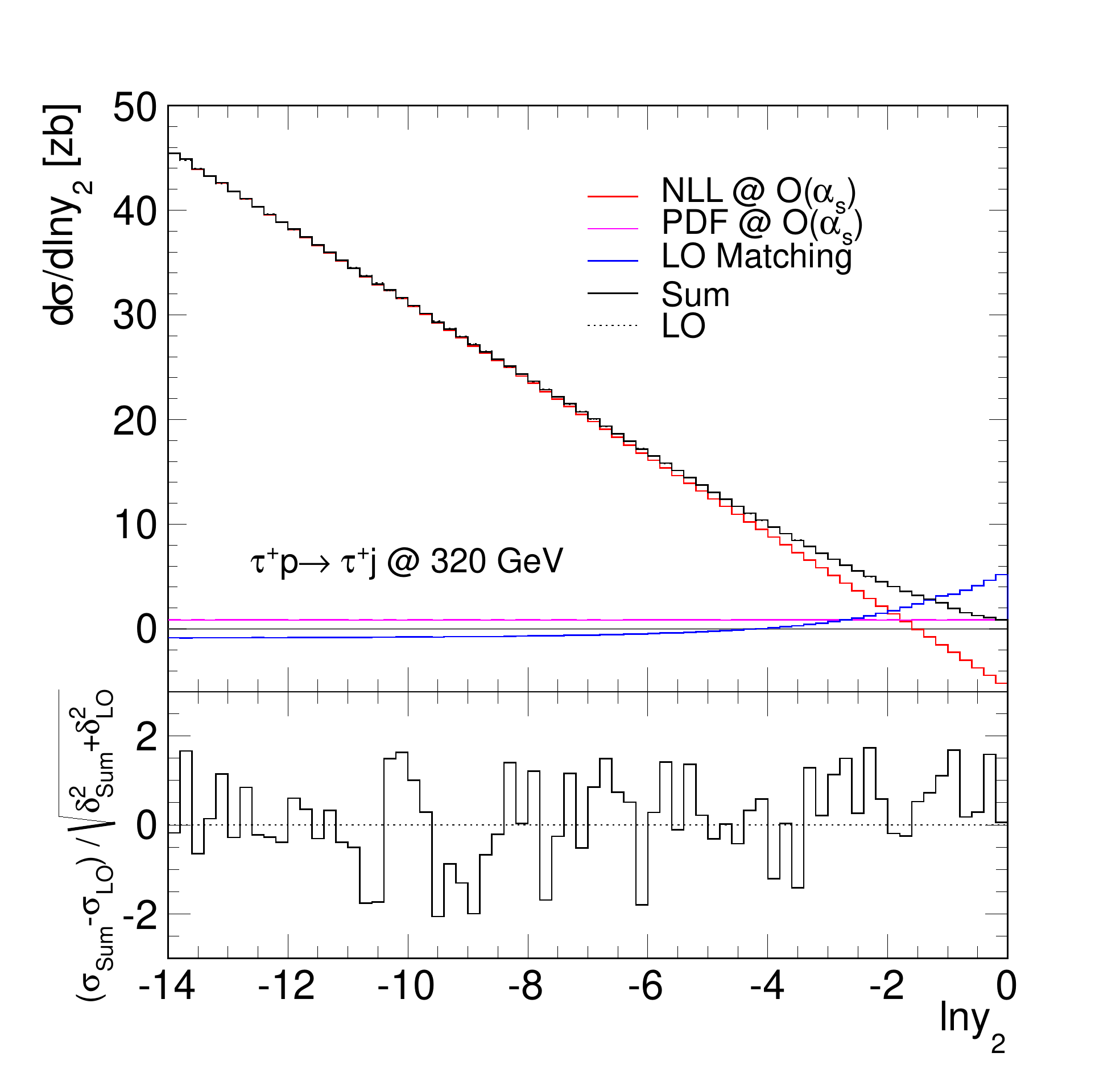}\\
    \includegraphics[width=0.42\textwidth]{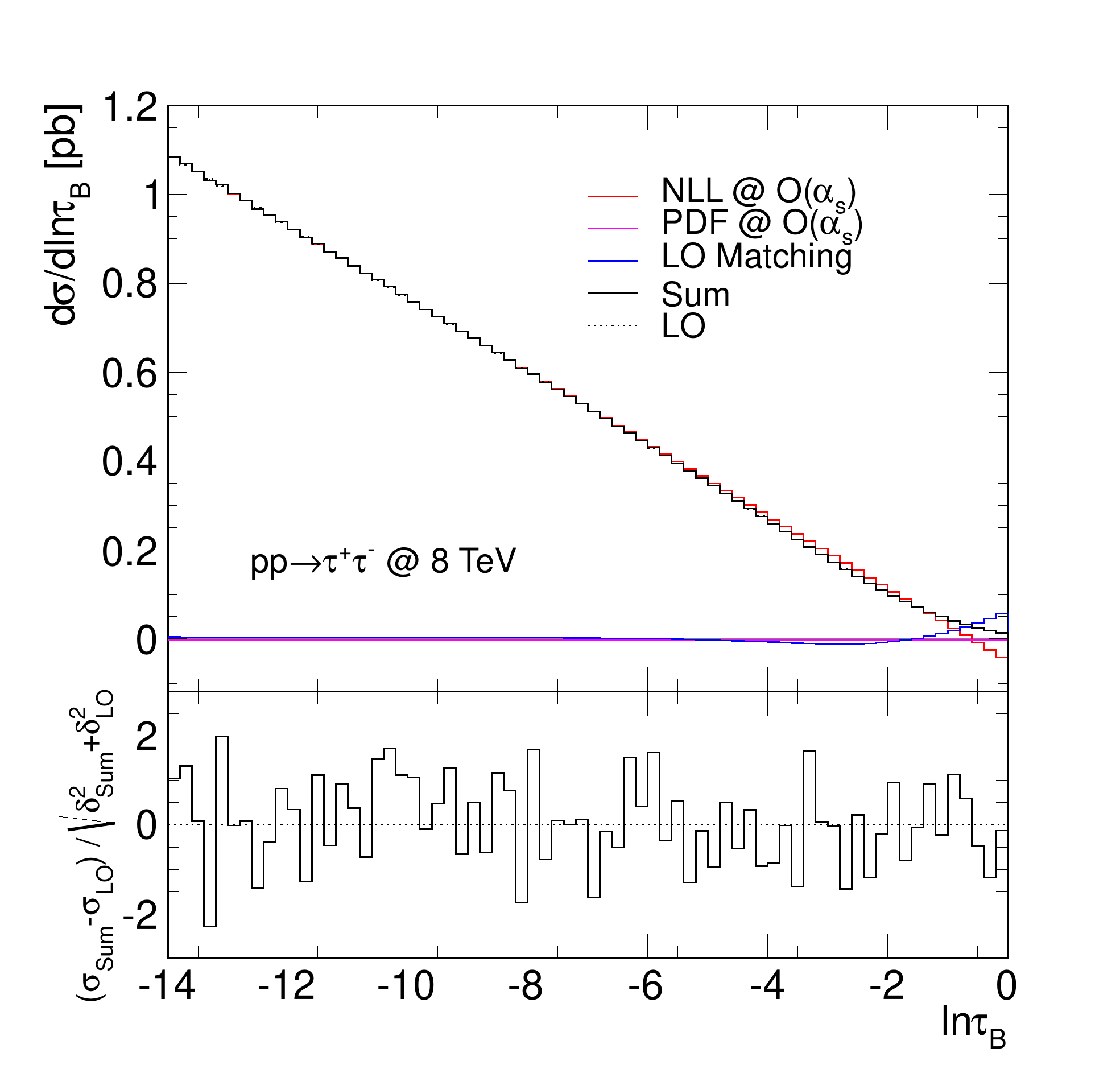}
    \includegraphics[width=0.42\textwidth]{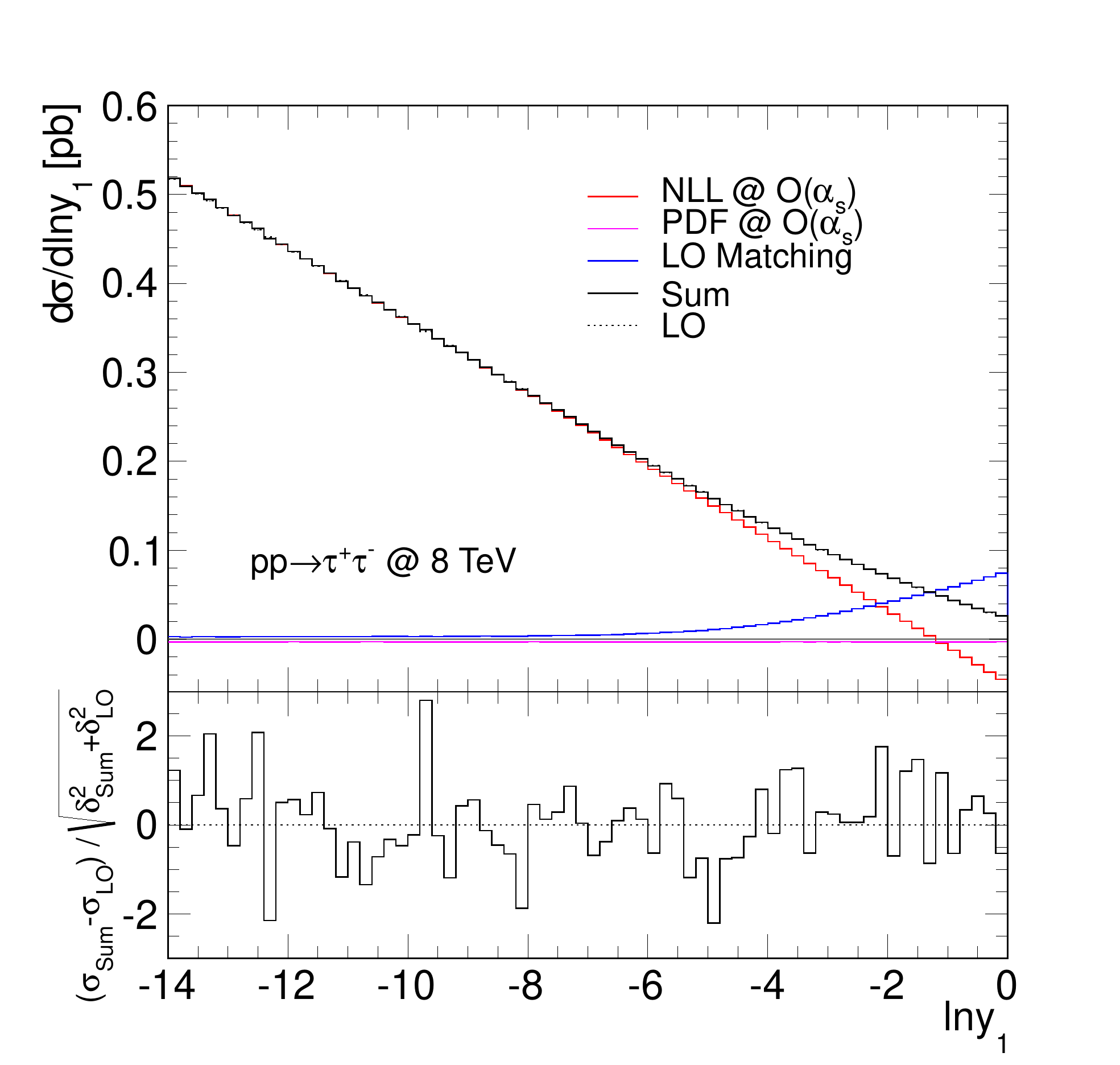}
  \caption{Test of the quasi-local matching procedure for hard processes with
    external gluons. Thrust (left panels) and leading jet rate (right panels)
    are compared between leading order and the first-order expanded 
    resummed and matched prediction for $\tau^+\tau^-\to gg$ (top),
    $\tau^+ g\to \tau^+g$ (middle) and $gg\to \tau^+\tau^-$ (bottom), all 
    mediated by Higgs-boson exchange.
    \label{fig:matching_gluons}}
\end{figure}

Figures~\ref{fig:matching_quarks} and~\ref{fig:matching_gluons} show in red the 
$\mathcal{O}(\alpha_s)$ expansion of the resummation, Eq.~\eqref{expansionLO_coeff}. We plot the 
result as a function of $\ln v$, using two observable types of different behavior with 
respect to the \caesar coefficients $a$ and $b_l$ ($a=b_l=1$ for thrust variables, on the 
left, while $a=2, b_l=0$ for jet rates, on the right. In both cases $d_l=g_l(\phi)=1$).  
The leading double logarithm appears as a straight line, while the sub-leading single 
logarithms appear as a constant offset. The collinear mass-factorization counterterms 
(the last term in the square bracket of Eq.~\eqref{expansionLO_coeff}) are shown in magenta, and 
the leading-order matching terms are displayed in blue. The sum of all the above is given 
in black. This sum is to be compared to a direct leading-order calculation, which is shown 
in black dashed. The difference between the two predictions should be of purely statistical 
nature, which is verified in the bottom panel of each plot by testing the relative size of 
the deviation, normalized to the Monte-Carlo uncertainty.


\subsection{A proof of concept: transverse thrust}\label{sec:global-event-shapes}
\begin{figure}
  \begin{center}
    \includegraphics[width=0.5\textwidth]{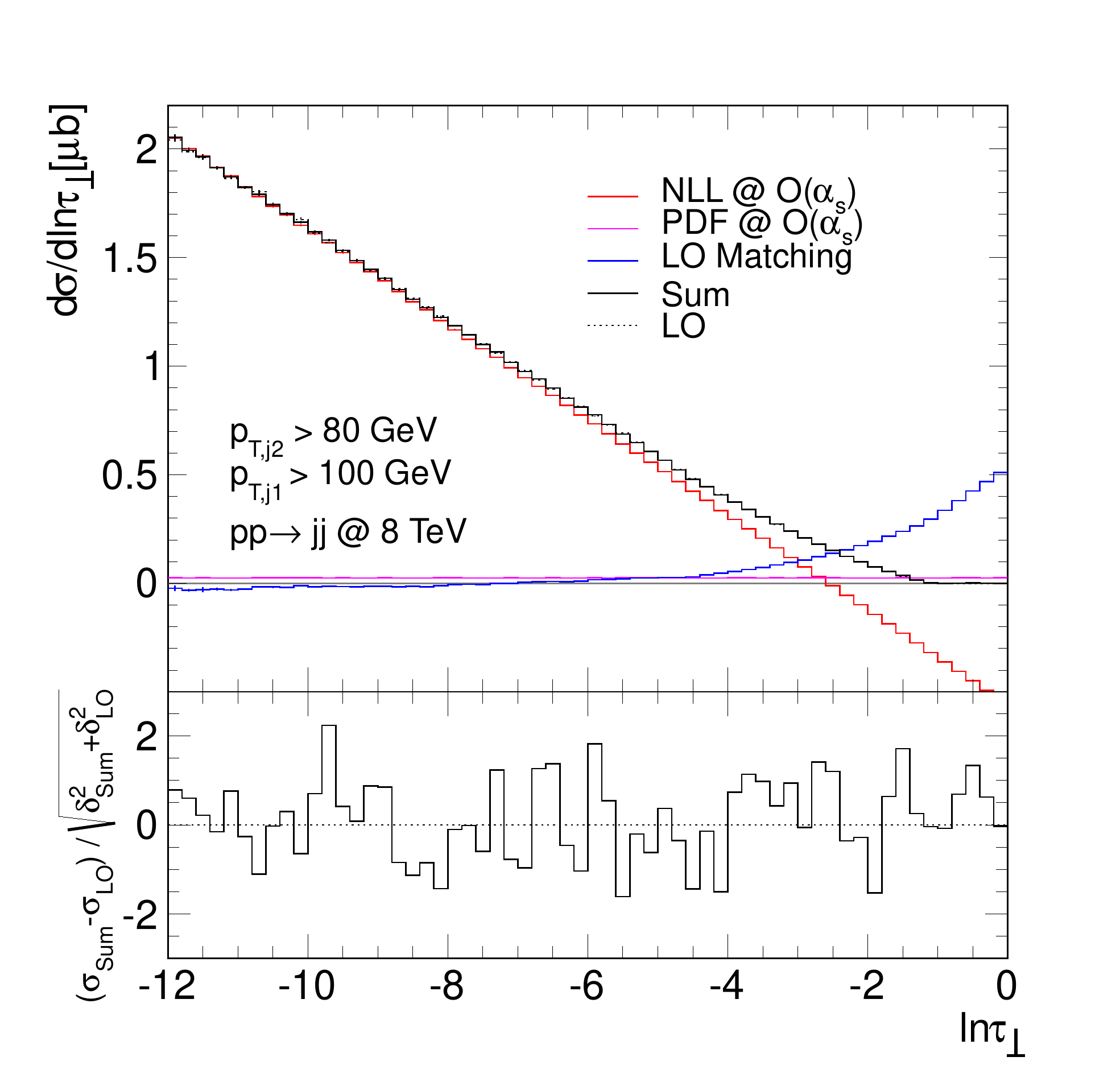}
  \end{center}
  \caption{Test of the matching procedure for transverse thrust.
    The leading-order prediction is compared to the first-order expanded 
    resummed and the LO matching term.
    \label{fig:matching_thrust}}
\end{figure}

\begin{figure}
  \begin{center}
  \includegraphics[width=0.46\textwidth]{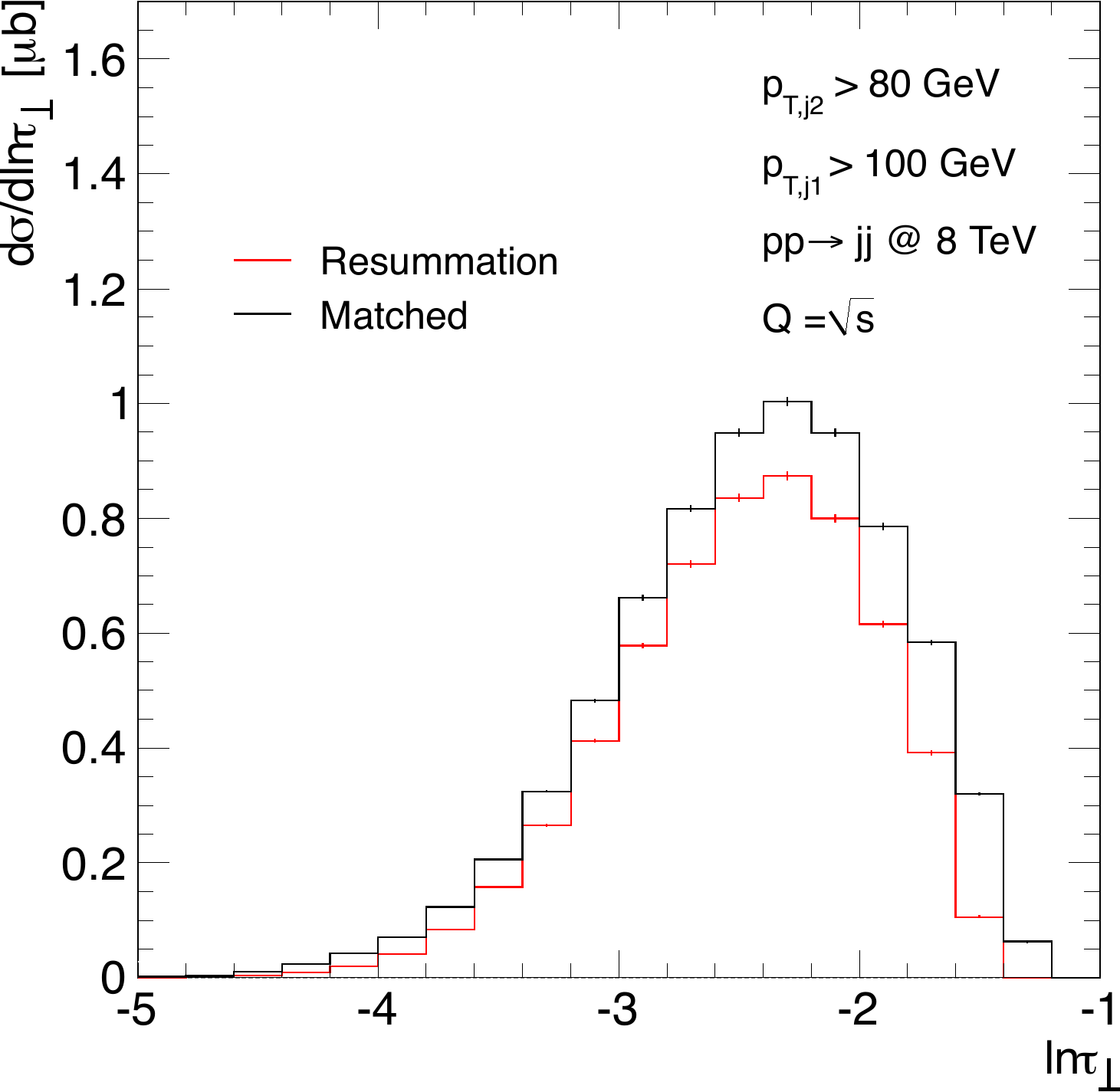}
   \includegraphics[width=0.46\textwidth]{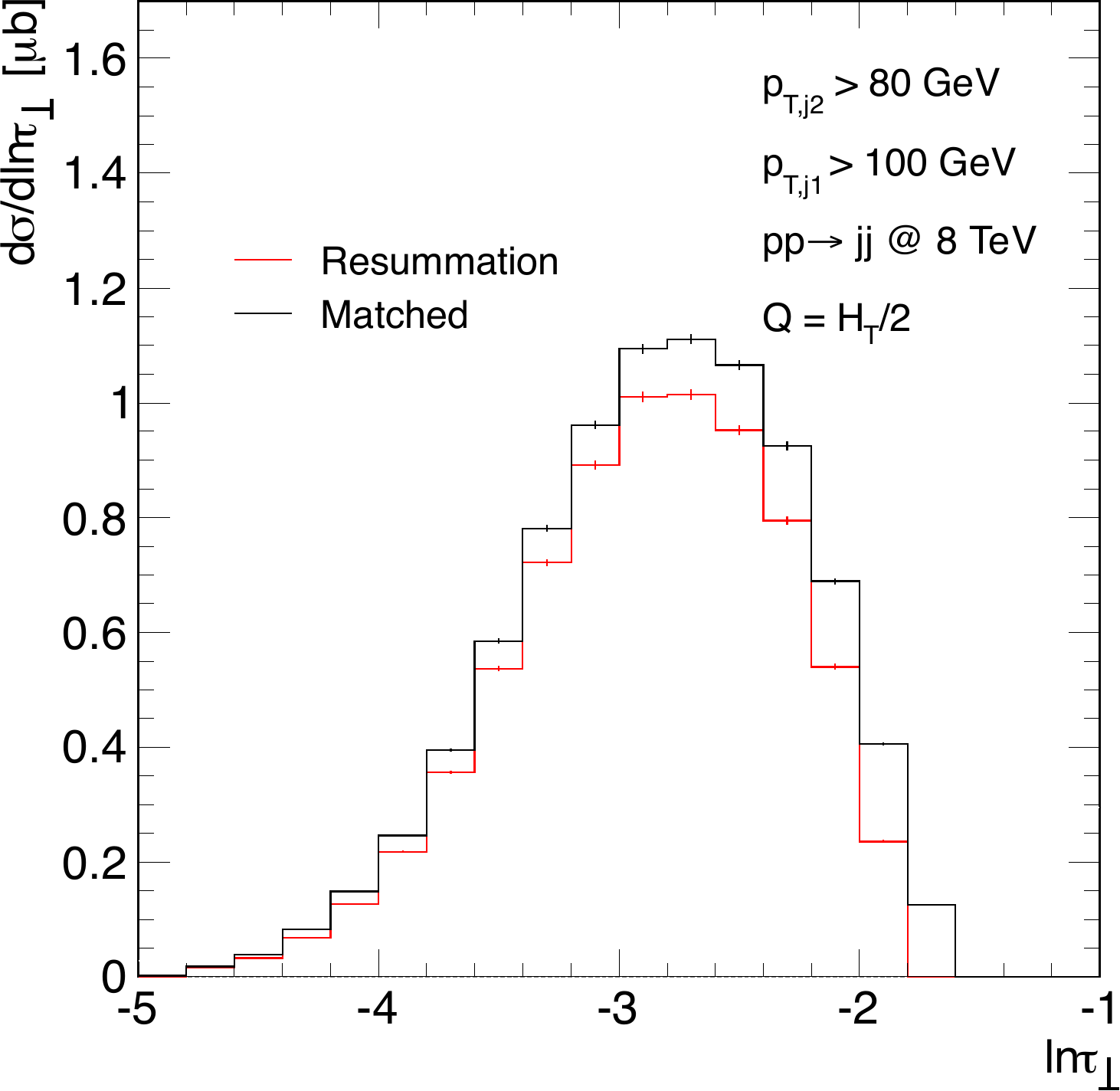}
     \end{center}
  \caption{The transverse-thrust distribution for $pp$ collisions at 8~TeV, with asymmetric cuts $p_{\perp 1}>100$~GeV, $p_{\perp 2}>80$~GeV.   \label{fig:transverse_thrust}}
\end{figure}

In order to demonstrate the completeness of our framework, we compute the resummed and matched distribution for a specific observable. We concentrate on the hadron-collider variant of the thrust observable, i.e. transverse thrust $T_\perp$. This global event-shape observable is defined as
\begin{equation}\label{transvthrust_def}
T_\perp=
\max_{\vec{n}_\perp}\frac{\sum_i |\vec{p}_{\perp i} \cdot \vec{n}_{\perp}  |}{\sum_i p_{\perp i}},
\end{equation}
where the sum runs over all final-state particles, with $\vec{p}_{\perp i}$  the particle's momentum 
transverse to the beam direction, and $p_{\perp i}=|\vec{p}_{\perp i}|$. The maximimal $T_\perp$ is found by variation of the transverse 
unit vector $\vec{n}_{\perp}$.
Transverse thrust has been studied by the Tevatron experiments~\cite{Bertram:2002sv, Aaltonen:2011et} and, more recently, also by the ATLAS~\cite{Aad:2012np} and CMS~\cite{Khachatryan:2014ika} collaborations.
Perturbative calculations for this distribution exist at NLO~\cite{Nagy:2003tz} and also at the resummed level in the \caesar framework~\cite{Banfi:2001bz,*Banfi:2003je,*Banfi:2004nk,*Banfi:2004yd}. In particular, the event shape that vanishes at Born level is $\tau_\perp=1-T_\perp$. Details of the resummation for a generic global event shapes are given in App.~\ref{app:caesar}. The response of the observable in the presence of soft / collinear emissions (see Eq.~(\ref{caesar-obs})) is parametrized by the coefficients given in Table~\ref{tab:transv-thrust}. Because the underlying Born processes is a $2\to 2$ QCD scattering, the color structure is non-trivial, hence we are able to put at work our construction of the soft function. In Figs.~\ref{fig:matching_thrust} and~\ref{fig:transverse_thrust}, we plot the transverse thrust distribution for $pp$ collisions at 8~TeV. We apply asymmetric cuts on the two leading jets, i.e. $p_{\perp 1}>100$~GeV, $p_{\perp 2}>80$~GeV and we set $\mu_R=\mu_F=H_T/2$, with $H_T=\sum_i p_{\perp i}$.

In particular, the plot in Fig.~\ref{fig:matching_thrust} is analogous to the ones already shown in Sec.~\ref{sec:automated-matching} and it provides yet another check of our matching procedure: the sum of the explicit expansion of the resummation (red), the collinear counterterm (magenta) and the LO matching term (blue) is plotted in solid black and it has to be compared to the LO calculation (dotted black). The bottom panel shows that the difference between the two is zero, within the Monte Carlo uncertainty. 

Finally, in Fig.~\ref{fig:transverse_thrust} we plot the resummed and matched distribution for transverse thrust (black curve). For comparison, we also show the resummation on its own (red curve). We show two possible choices for the hard scale: $Q=\sqrt{s}$ (on the left) and $Q=H_T/2$ (on the right). The latter, more natural in hadron-hadron collisions, corresponds to a rescaling of the resummation coefficients in Table~\ref{tab:transv-thrust}, namely $d_l \to d_l \left(H_T/ 2 / \sqrt{s} \right)^a$.The key feature of this plot is that the soft function and  matching are computed in a fully automated way at run-time, leading to NLL resummed and matched 
distributions with a similar level of automation as Monte Carlo event generators.

%Nevertheless, the plot in Fig.~\ref{fig:transverse_thrust} demonstrates that we have successfully construct a flexible %framework where these studies are possible.

\begin{table}[t]
  \begin{center}
%  \begin{tabular}{c|c|c|c|c|}
    \begin{tabular}{c|cccc}
  leg $l$ & $a_l$ & $b_l$ & $d_l$&$g_l(\phi)$ \\
 \midrule
  $1$ &   $1$ & $0$ & $1/\sin \theta$ & $1-|\cos \phi|$ \\
  $2$ &   $1$ & $0$ & $1/\sin \theta$ & $1-|\cos \phi|$ \\
  $3$ &   $1$ & $1$ & $1/\sin^2 \theta$ & $\sin^2 \phi$ \\
  $4$ &   $1$ & $1$ & $1/\sin^2 \theta$ & $\sin^2 \phi$
   \end{tabular}
  \end{center}
  \caption{Coefficients of the \caesar formula that specify the NLL resummation of transverse thrust~\cite{Banfi:2001bz,*Banfi:2003je,*Banfi:2004nk,*Banfi:2004yd}. They correspond to the choice for the hard scale $Q=\sqrt{s}$; $\theta$ being the scattering angle in the partonic centre of mass frame, and $\phi$ denoting the azimuth.}
  \label{tab:transv-thrust}
\end{table}

%% file: inputs/summary.tex
\section{Conclusions and Outlook}\label{sec:conclusions}

Multi-jet physics is central in the physics program of the LHC. In this paper, we have overcome the two main technical difficulties that prevented NLL resummed calculations to be performed in processes with high jet multiplicity. 

The first issue was related to the color structure of soft emissions at wide angle, i.e.
 away from the jets, the complexity of which rapidly increases with the number of hard jets. We have solved this problem by constructing and implementing a framework in which the NLL soft function is computed in an highly automated way.
The algorithm constructs an appropriate color basis for the partonic process at hand, and evaluates color operators and the decomposition of Born amplitudes in this basis. It makes use of the matrix-element generator \comix to access the color-ordered partial amplitudes that are needed for the evaluation of the soft function. 
Using this framework, we have obtained and validated results for the soft function for all QCD processes with up to five hard jets in the final state, i.e.\ $2 \to 5$ QCD amplitudes, and we have studied the validity of the widely used large $\nc$ approximation. We have found that the impact of finite-$\nc$ corrections is significant, especially for processes with many gluons. 

We have tackled the second problem of matching resummed predictions to fixed-order calculations. In the traditional way of addressing this problem, one matches the resummed distribution of a given observable $v$ to the one obtained at fixed-order (typically NLO). The main drawback of this approach is that the fixed-order result and the expanded resummed result have to be computed independently in extreme regions of phase space, i.e.\ at very small $v$, where numerical cancellation is hard to achieve.
We improve upon this situation by introducing a quasi-local matching scheme at leading order, which generates the finite remainder directly.
As a proof of concept, we computed within our framework the NLL transverse-thrust distribution matched to LO. Although in this study we have mainly concentrated on global event shapes, our framework can be easily extended to the case of non-global observables.

We see this rather technical paper as the first necessary step in a rich program aimed  
at the phenomenological applications of resummed perturbation theory in multi-jet physics.
Moreover, because we implement resummed calculations in the \sherpa framework, we have the possibility of making precise comparisons between analytic resummation and Monte-Carlo parton showers. This will provide insights on the benefits and limitations of both approaches and perhaps even indicate ways to improve the formal accuracy of the parton shower.

%% file: inputs/appendix.tex
\section{Over-complete color bases}
\label{sec:app_bases}

Color bases constructed from irreducible QCD representations do 
not meet our requirement \ref{req_minimal_partial_count} of the list in Sec.~\ref{sec:color-flow}.  
Although we do not automate the construction of orthonormal bases 
this for work, we do employ their generic properties in several 
arguments throughout this section.  
For a complete approach to their construction see~\cite{Keppeler:2012ih}.

We define an orthogonal basis element $e_{\alpha}$ so that 
\begin{equation}
\langle e_\be | e_\al \rangle = e_{\be\al} = \lambda_\al 
\delta_{\be\al}\,, \qquad \text{(no sum on $\alpha$)}\,
\label{orth_bas}
\end{equation}
where $\lambda_\alpha$ is the weight of the representation.
For a given physical process, the dimensionality of the 
orthogonal $e$-basis may differ from the $c$-basis, in which case 
the indices in Eq.~\eq{north_bas} versus Eq.~\eq{orth_bas} also differ.  
Starting with the metric $e_{\alpha \beta}$, we define 
the (possibly non-square) transformation to the $c$-basis via
\begin{alignat}{5}
R : \to R^{\;\;\al'}_{\al} e_{\al'\be'} (R^T)^{\be'}_{\;\;\be} & = 
&\; c_{\al\be}
\qquad
 R_{\;\;\al'}^{\al} e^{\al'\be'} (R^T)_{\be'}^{\;\;\be} & = 
&\; c^{\al\be} .
\label{Rnor}
\end{alignat}
As both $c_{\al\be}$ and $c^{\al\be}$ are symmetric, their 
(independent) eigenvectors correspond to the row elements 
of $R$ providing a straight-forward construction.

\subsection[General proofs for the $N_C = 3 + \epsilon$ expansion]{General proofs for the \boldmath{$N_C = 3 + \epsilon$} expansion}\label{sec:NCeps_proof}

\subsubsection*{Lemma 1} 

Assuming $\nc = 3+\epsilon$, with $\epsilon>0$ and $\epsilon \ll 1$,  we can cleanly separate the finite part of the inverse metric 
from the divergent one, i.e. 
\begin{equation} \label{eq:app-inv-metr}
c^{\al\be} |_{\nc = 3+\epsilon} = 
\frac{1}{\epsilon} \tilde{c}^{\,\al\be} + c^{\al\be}_R  |_{\nc = 3}+ {\cal O}(\epsilon),
\end{equation}
where $c^{\al\be}_R$ and $ \tilde{c}^{\,\al\be}$ are 
regular at $\nc =3$. 

\subsubsection*{Proof} 

For general $\nc$, the metric can be brought to 
diagonal form where the $\lambda_\alpha$ are polynomial in $\nc$ corresponding 
to the weights of irreducible representations, which in the limit 
$\epsilon \to 0$ are either $\ope{(1)}$ or $\ope{(\epsilon)}$. In the latter case,
let us parameterize such eigenvalues as $\lambda_0 = \kappa \epsilon$ where $\kappa$ is a constant\footnote{The case $\lambda_\alpha |_{\nc=3} = \ope{(\epsilon^2)}$ 
 is in principle possible and it would lead to an $\ope{(\frac{1}{\epsilon^2})}$ term in Eq.~\eq{eq:app-inv-metr}. However, this situation has not been encountered for any of the color-flow bases considered in this study.}. 
 
 The inverse of the orthogonal 
metric is a matrix with diagonal entries $1/\lambda_\alpha$.  
We construct the tensor $\tilde{c}^{\,\al\be}$ by rotating only the 
$ 1/\lambda_0$ components back to the $c$-basis.  Defining 
$\alpha'_0$ as the indices running over the vanishing weights 
we have 
\begin{equation}
 \tilde{c}^{\,\al\be} =  \kappa \,
R^{\al}_{\;\;\al'_0} \delta^{\al_0'\be'} \left(R^T\right)^{\;\;\be}_{\be'} .
\end{equation}

\subsubsection*{Lemma 2}

All color products $\T_i\,\cdot\,\T_j$ belong to the null space of the singular part of the inverse metric, i.e.
\begin{equation}
\tilde{c}^{\,\al\be} (\T_i\,\cdot\,\T_j)_{\beta \gamma} = {\mathbf 0}^{\al}_{\;\;\gamma}\,.
\label{vans3}
\end{equation} 

\subsubsection*{Corollary}
An interesting, and computationally advantageous, consequence of the above Lemma is that in order to obtain
\begin{alignat}{5}
 \mathcal{S}(\xi)_{\nc =3 + \epsilon}  = \mathcal{S}(\xi)_{\nc = 3} +  \ope{(\epsilon)}\,. 
 \label{keyr}
\end{alignat} 
we only have to evaluate the color metric and its inverse with $\nc = 3+ \epsilon$, while computing all color producs $\T_i \,\cdot \,\T_j$ at $\nc=3$. 
Therefore, the inversion of the metric at $\nc=3+\epsilon$ (with $\epsilon$ small) provides a valid alternative to dimensional reduction for computing the soft function\footnote{A difficulty arises in 
our method for the case of $6$ gluon soft 
evolution in the trace basis.  The problem is linked to the fact 
that $9$ of the vanishing $\nc =3$ eigenvalues are negative 
for $3 <\nc \lesssim 3.32$.  Therefore, an inversion algorithm for 
$\nc=3+\epsilon$ 
dependent on positive definiteness of the (symmetric) metric is 
incompatible.  However, this problem is avoided by 
choosing the $f$-basis, which is positive definite for all processes 
considered thus far, or inverting using a (much slower) more 
generic algorithm.  A similar problem arises in the standard basis for $qq\to qqggg$.}.

\subsubsection*{Proof}

Rotating $\tilde{c}^{\; \al\be}$ to the e-basis we define an element
\begin{equation}
\langle e_\alpha |  =  \langle c_{\alpha'} | R^{\alpha'}_{\;\;\alpha}\,,
\end{equation}
so that for every $0$ eigenvalue of $c_{\alpha \beta}$ 
there is a corresponding element $\langle e_{\alpha_0} |$ 
in the orthogonal basis which satisfies
\begin{equation}
\langle e_{\alpha_0} | e_{\beta} \rangle =  0 \quad \forall \beta \,.
\end{equation}
Since $\T_i \,\cdot \,\T_j | e_{\beta} \rangle$ is a coefficient 
times an element of the orthogonal basis, we conclude that 
$(\T_i \cdot \T_j)_{\alpha_0 \beta} = 0$.  Repeating the 
argument and noting that $c^{\al\be}$ is symmetric, we 
conclude that the corresponding rows and columns of 
$(\T_i \cdot \T_j)$ are zero.  In the $e$-basis we clearly 
have
\begin{equation}
\delta^{\al \al_0 } \langle e_{\al_0} |\, \T_i \cdot \T_j \,| e_{\gamma} \rangle \; 
\Rightarrow \;
\tilde{c}^{\,\al\be} (\T_i \cdot \T_j)_{\beta \gamma} = {\mathbf 0}^{\al}_{\;\;\gamma}\,,
\end{equation}
which gives the desired result. \medskip

In order to demonstrate the corollary we evaluate all color products at $\nc=3+\epsilon$ and we then write the soft anomalous dimension Eq.~(\ref{eq:gamma-ind}) at small $\epsilon$ as
\begin{equation}
 \Gamma_{\alpha \beta}|_{\nc = 3 + \epsilon}
 = 
 \Gamma_{\alpha \beta}|_{\nc = 3 } 
 +
 \epsilon \, \tilde{\Gamma}_{\alpha \beta}|_{\nc = 3 }\,. 
\end{equation} 
The first term contributing to the soft function $\mathcal{S}$ which 
involves the inverse metric comes from expanding the exponential 
to second order
\begin{equation}
 \mathcal{S}(\xi) \sim \frac{\xi^2}{2!}\left[ \Gamma_{\beta \al}
 +
 \epsilon \, \tilde{\Gamma}_{\beta \al} \right]
 	\left[c^{\alpha \gamma}_R  + \frac{1}{\epsilon} \tilde{c}^{\,\alpha \gamma} \right]
 \left[ \Gamma_{\gamma \alpha'}
 +
 \epsilon \, \tilde{\Gamma}_{\gamma \alpha'} \right].
\end{equation} 
Using \eq{vans3} on all the color products that enter the definition of $\Gamma$, one finds no finite terms originating from the interference of the $1/ \epsilon$ pole of the inverse metric and the $\ope{(\epsilon)}$ contribution to the anomalous dimensions. Furthermore, this holds for higher terms in the expansion of 
the exponential. Therefore, all the color products necessary to construct the soft anomalous dimension can be safely computed at $\nc=3$, while it is still necessary to compute the metric and its inverse at $\nc=3+\epsilon$

Finally, we note that using $\nc = 3 +\epsilon$ to invert the metric involves a  
large cancellation among the entries of $c^{\alpha\beta}$.  However, the convergence 
is better than expected since the coefficients of $1/\epsilon$ are roughly proportional 
to the number of corresponding ${\bf 0}$-representations, which is 
smaller than the total number of irreducible representations for a given process.  
\smallskip

\subsection[A concrete example: $gg\to gg$]{A concrete example: \boldmath{$gg\to gg$}}
\label{sec:gggg_ex}

We list here several different manifestations of the 4-gluon 
basis as specific examples for our more general discussion 
in the text.

\subsubsection*{Trace basis}
Let us consider the trace basis for this process:
\begin{alignat}{5} \label{gggg_over}
c_1 &= K_c (\tf_{a_{1}}\tf_{a_{2}}\tf_{a_{3}}\tf_{a_{4}} + \tf_{a_{1}}\tf_{a_{4}}\tf_{a_{3}}\tf_{a_{2}}),
\qquad c_4 &= K_d \delta_{a_1 a_2} \delta_{a_3 a_4},
\notag \\
c_2 &= K_c (\tf_{a_{1}}\tf_{a_{2}}\tf_{a_{4}}\tf_{a_{3}} + \tf_{a_{1}}\tf_{a_{3}}\tf_{a_{4}}\tf_{a_{2}}),
\qquad c_5 &= K_d \delta_{a_1 a_3} \delta_{a_2 a_4},
\notag \\
c_3 &= K_c (\tf_{a_{1}}\tf_{a_{3}}\tf_{a_{2}}\tf_{a_{4}} + \tf_{a_{1}}\tf_{a_{4}}\tf_{a_{2}}\tf_{a_{3}}),
\qquad c_6 &= K_d \delta_{a_1 a_4} \delta_{a_2 a_3},
\end{alignat}
where $\tf_{a_{i}}$ are the color generators in the fundamental representation and a trace over their fundamental-representation indices is implicit. 
Note that $K_c = \nc( 16 \nc^6 - 3 \nc^4 +16 \nc^2 - 6)^{-\frac{1}{2}}$ and 
$K_d = (\nc^2-1)^{-1}$, so that the basis is normalized.
The tree-level partial amplitude coefficients corresponding to these 
color basis elements are 
\begin{alignat}{5}
\label{hard_t_gggg}
c_1 \to m_0(1,2,3,4), \quad
c_2 \to m_0(1,2,4,3), \quad
c_3 \to m_0(1,4,2,3), \quad
c_4 = c_5 = c_6 = 0.
\end{alignat}

\subsubsection*{Dimensionally reduced trace basis}

The basis in Eq.~(\ref{gggg_over}) is over-complete and consequently the color metric is not invertible for $\nc=3$. However, if we consider a reduced basis, obtained by taking
\begin{alignat}{5}
c'_1 & = c_1, \qquad
c'_2 & = c_2, \qquad 
c'_3 & = c_3, \qquad
c'_4 & = \frac{K_d' } {K_d} (c_4 + c_6), \qquad
c'_5 & = \frac{K_d' } {K_d} (c_5 + c_6) ,
\end{alignat}
where $K_d' = 1/\sqrt{2\nc^2(\nc^2-1)}$, the metric is then invertible for 
$\nc =3$, the connected components remain synced with the 
hard matrix \eq{hard_t_gggg}, and the resulting $\mathcal{S}(\nc = 3)$ is unchanged.  

\subsubsection*{Adjoint basis}

We can now write the 5-dimensional $f$-basis
\begin{alignat}{5}
c_1 &= K_a f^{a_1 a_4 e_1} f^{e_1 a_3 a_2},
\qquad
&& c_2 = K_a  f^{a_1 a_3 e_1} f^{e_1 a_4 a_2},
\notag \\
c_3 &= K_d \delta_{a_1 a_4} \delta_{a_2 a_3}, 
&& c_4 = K_d \delta_{a_1 a_2} \delta_{a_3 a_4}, \qquad
c_5 = K_d \delta_{a_1 a_3} \delta_{a_2 a_4},
\label{bggggf}
\end{alignat}
where $K_a = (4 \nc^2 (\nc^2-1))^{-\frac{1}{2}}$ and 
$K_d = (\nc^2-1)^{-1}$.  We can make connection to 
the trace basis by repeated application of the fundamental  
Lie algebra to see that
\begin{alignat}{5}
c_1 &= K_a \left [(\tf_{a_{1}}\tf_{a_{2}}\tf_{a_{4}}\tf_{a_{3}} + \tf_{a_{1}}\tf_{a_{3}}\tf_{a_{4}}\tf_{a_{2}})
-(\tf_{a_{1}}\tf_{a_{2}}\tf_{a_{3}}\tf_{a_{4}} + \tf_{a_{1}}\tf_{a_{4}}\tf_{a_{3}}\tf_{a_{2}})\right],
\notag \\
c_2 &= K_a \left[(\tf_{a_{1}}\tf_{a_{2}}\tf_{a_{3}}\tf_{a_{4}} + \tf_{a_{1}}\tf_{a_{4}}\tf_{a_{3}}\tf_{a_{2}})
-(\tf_{a_{1}}\tf_{a_{4}}\tf_{a_{2}}\tf_{a_{3}} + \tf_{a_{1}}\tf_{a_{3}}\tf_{a_{2}}\tf_{a_{4}})\right].
\label{bggggf2}
\end{alignat}
in terms of fundamental representation generators.  The hard coefficients are the 
same partial ordered amplitudes though we now have multi-peripheral labelling 
\begin{alignat}{5}
\label{hard_adj}
c_1 \to m_0(1,3,4,2), \quad
c_2 \to m_0(1,4,3,2), \quad
c_3 = c_4 = c_5 = 0.
\end{alignat}
The evaluation of $\mathcal{S}(\xi)$ in the adjoint basis is equivalent 
to the trace basis at $\nc = 3$.

\subsubsection*{Inversion with \boldmath{$N_C = 3 + \epsilon$}}

We consider here the trace basis for $\nc = 3 + \epsilon$. We note that there are no additional null eigenvalues for $\nc \ne 3$. We then take the $\epsilon \to 0$ limit and expand the matrix representing the color metric in terms of its regular and singular pieces, as in Eq.~(\ref{eq:app-inv-metr}). The residue of the $1/ \epsilon$ pole is
\begin{alignat}{5}\label{mate_no}
\tilde{c}^{\alpha\beta} \; = \;
\left(
\begin{array}{ccc}
 K_{3\times 3}(\frac{23}{27} )& K_{3\times 3} (\sqrt{\frac{92}{243}}) \\
 K_{3\times 3} (\sqrt{\frac{92}{243}}) & K_{3\times 3}(\frac{4}{9})   \\
\end{array}
\right),
\end{alignat}
where $K_{3\times 3}(a)$ is a $3\times 3$ matrix with each element 
equal to $a$.  The matrix in Eq.~\eq{mate_no} is precisely the 
trace-basis form for the inverse eigenvalue 
$\lambda_0 \sim 1/\epsilon$ of the $\nc = 3$ $0$-representation in 
the orthogonal basis.  One can verify that the matrix $(\T_i \cdot \T_j)_{\alpha \beta}$ 
at $\nc = 3 $ for all $i$ and $j$ is in the null space of $\tilde{c}^{\alpha\beta}$. This a concrete manifestation of the behavior expected from the discussion in App.~\ref{sec:NCeps_proof}.  

\subsection{Bases properties for multi-parton processes}\label{app:mp_bases}
In Tables~\ref{bases_45} and \ref{bases_67} we summarize the main properties of the multi-parton processes considered in Sec.~\ref{sec:resum}.

\begin{table}[h!]
\begin{center}
     \begin{tabular}{||c||c|c|c|c|c|c||}
		\hline
    Sub-process   & $gggg$  & $q\bar{q}gg$  & $q\bar{q}q\bar{q}$ & $ggggg$ & $q\bar{q}ggg$ & $q\bar{q}q\bar{q}g$ \\ \hline
    		\hline
    Dim. basis   & 5  & 3  & 2 & 16 & 10 & 4 \\ \hline
    			\hline
    Dim. Born   & 2 & 2  & 2 & 6 & 6 & 4 \\ \hline
    			\hline
     zero eigenvalues & 0  & 0  & 0 & 0 & 0 & 0 \\ \hline
    \end{tabular}
\end{center}
\caption{ Summary of basis properties for all $4$- and $5$-parton processes.  
Pure gluon processes are listed in the adjoint $f$-basis.}
    \label{bases_45}
\end{table} 

\begin{table}[h!]
\begin{center}
     \begin{tabular}{||c||c|c|c|c|c|c|c|c||}
		\hline
    Sub-process   & $6g$  & $q\bar{q}4g$  & $q\bar{q}q\bar{q}gg$ & $qqqq\bar{q}q$ & $7g$ 
    & $q\bar{q}5g$ & $q\bar{q}q\bar{q}3g$ & $q\bar{q}q\bar{q}q\bar{q} g$ \\ \hline
    		\hline
    Dim. basis   & 79  & 46  & 14 & 6 & 421 & 252  & 62 & 18\\ \hline
    			\hline
    Dim. Born   & 24  & 24  & 12 & 6 & 120 &  120 & 48 & 18 \\ \hline
    			\hline
     zero eigenvalues  & 5  & 6  & 1 & 0 & 70 & 75 & 12 & 1\\ \hline
    \end{tabular}
\end{center}
\caption{Summary of basis properties for all $6$- and $7$-parton processes.  Pure 
gluon processes are listed in the adjoint $f$-basis.}
    \label{bases_67}
\end{table} 

\newpage
\section{The \caesar framework}\label{app:caesar}
\caesar~\cite{Banfi:2001bz,*Banfi:2003je,*Banfi:2004nk,*Banfi:2004yd} is a computer program that allows one to perform the resummation of a large class of observables, namely global event shapes, to NLL accuracy. In this appendix we recap, without re-deriving them, the expressions of the leading and next-to-leading function $g_1^{(\delta)}$ and $g_2^{(\delta,\mathcal{B})}$ in Eq.~(\ref{eq:fact}) as obtained in the \caesar framework.
The LL function reads
\begin{align}\label{caesar-g1}
g_1^{(\delta)}(\as L)&=\nonumber \\ &-\sum_{l=1}^n  \frac{C_l}{2 \pi \beta_0\lambda b_l}\left[ 
(a-2 \lambda) \ln \left( 1-\frac{2\lambda}{a}\right)-(a+b_l-2 \lambda)\ln \left(1-\frac{2 \lambda}{a+b_l}\right)\right],
\end{align}
where $\lambda=\as \beta_0 L$, $\as=\as(\mu_R^2)$ and $\beta_0$ is the one-loop coefficient of the QCD $\beta$-function, $\beta(\as)= -\as \left(\as \beta_0+\as^2 \, \beta_1+\dots\right)$, with
\begin{equation}\label{QCD-beta0}
 \beta_0=\frac{11 C_A-2 n_f}{12 \pi},\quad \beta_1=\frac{17 C_A^2-5 C_A n_f-3 C_F n_f}{24 \pi^2}.
\end{equation}
The result in Eq.~(\ref{caesar-g1}) consists of a sum over all the hard partons and the dependence on the color is trivial and only enters through the Casimir of each leg $l$, ($C_F$ for a quark leg, $C_A$ for a gluon leg).
Note also that $a_1=a_2=\dots=a_n=a>0$.

The result for the NLL function $g_2^{(\delta,\mathcal{B})}$ has a richer structure:
\begin{align}\label{g2-caesar}
g_2^{(\delta,\mathcal{B})}(\as L)&= -\sum_{l=1}^n C_l \left[ \frac{r^{(2)}_l}{b_l}+ B_l \, T \left( \frac{L}{a+b_l} \right) \right]+\partial_L \left[ Lg_1^{(\delta)}(\as L)\right] \left(\ln \bar{d}_l -b_l \ln \frac{2 E_l}{Q}\right)
\nonumber\\
&+ \sum_{l=1}^{n_\text{initial}} \ln \frac{q^{(l)}(x_l, \mu_F^2e^{-\frac{2 L}{a+b_l}})}{q^{(l)}(x_l, \mu_F^2)}+\ln \mathcal{F}\left(\partial_L Lg_1^{(\delta)}(\as L) \right)\nonumber \\&-T\left(L/a\right)\sum_{l=1}^n C_l \ln\frac{Q_{12}}{Q}+ \ln \mathcal{S}\left(T\left(L/a\right)\right).
\end{align}
The first term in the square brackets in Eq.~(\ref{g2-caesar}) contains the two-loop contributions to the DGLAP splitting function in the soft limit and to the QCD $\beta$-function, as well as the dependence on the renormalization scale $\mu_R$:
\begin{align}\label{caesar-g2-r2}
r_l^{(2)}&=\left( \frac{K}{4 \pi^2 \beta_0^2}-\frac{1}{2\pi \beta_0} \ln\frac{\mu_R^2}{Q^2}\right)\left[ \left( a+b_l\right) \ln \left(1-\frac{2 \lambda}{a+b_l}\right)-a \ln\left(1-\frac{2 \lambda}{a}\right) \right] \nonumber \\
&+\frac{\beta_1}{2 \pi \beta_0^3}\left[ \frac{a}{2} \ln^2 \left(1-\frac{2\lambda}{a}\right)-\frac{a+b_l}{2}\ln^2\left(1-\frac{2 \lambda}{a+b_l}\right) +a \ln\left(1-\frac{2 \lambda}{a}\right) \right.
\nonumber \\ & \left.
-(a+b_l)\ln\left(1-\frac{2 \lambda}{a+b_l}\right)\right], \quad \text{with} \quad K=C_A\left(\frac{67}{18}-\frac{\pi^2}{6}\right)-\frac{5}{9}n_f.
\end{align}
The second term in the square brackets instead captures hard collinear emissions to a quark leg ($B_q=-\frac{3}{4}$) or to a gluon leg ($B_g=-\pi \beta_0$); we have introduced
\begin{equation}\label{Tfunction}
T(L) = \frac{1}{\pi \beta_0} \ln\frac{1}{1-2 \alpha_s \beta_0 L}
\end{equation}
The last term of the first line of Eq.~(\ref{g2-caesar}) contains
\begin{equation}\label{logdbar}
\ln \bar{d}_l = \ln d_l +\int_0^{2\pi }\frac{d \phi}{2 \pi} \ln g_l(\phi),
\end{equation}
while $E_l$ is the energy of leg $l$. We note that the contribution in this round brackets is actually frame-independent. We move then to the second line of Eq.~(\ref{g2-caesar}) and the first term we encounter is the one that depends on the PDFs ($\mu_F$ is the factorization scale). This contribution comes about because we veto emissions collinear to the incoming legs which would contribute to the event shape more than a quantity $v$.  There is then a term ($\mathcal{F}$) describing the effect of multiple emissions. The calculation of this term is highly non-trivial for generic observables and indeed this is one of the central aspects of the analysis of Refs.~\cite{Banfi:2001bz,*Banfi:2003je,*Banfi:2004nk,*Banfi:2004yd}. However, at NLL, multiple emissions have a color structure identical to $g_1^{(\delta)}$, thus this term is trivial from the point of view of our current analysis. For additive observables, like, for instance, transverse thrust considered in Sec.~\ref{sec:global-event-shapes}, this multiple-emission contribution has a rather simple form
\begin{equation}\label{t-thrust-mult-em}
\mathcal{F}(L) = \frac{e^{\gamma_E \partial_L \left( L g_1^{(\delta)}(\as L)\right)}}{\Gamma\left(1-\partial_L\left( L g_1^{(\delta)}(\as L) \right)\right)}.
\end{equation}
Finally, in the last line we encounter the contributions due to soft radiation at large angle, which we can, for convenience, divide into a diagonal contribution and one with a non-trivial matrix structure.

Thus, all the terms but the last one in the \caesar master formula Eq.~(\ref{g2-caesar}) are diagonal in color and therefore apply to processes with an arbitrary number of hard legs. The results of Sec.~\ref{sec:resum} provide an automated way of computing the only contribution at NLL with a non-trivial color structure, namely the soft function $\mathcal{S}$, which captures the effect of soft gluon emitted at wide angles.